\documentclass[graybox, natbib]{svmult}
\bibpunct{(}{)}{;}{a}{}{,} 

\usepackage{mathptmx}       
\usepackage{helvet}         
\usepackage{courier}        
\usepackage{type1cm}        

\usepackage{makeidx}         
\usepackage{graphicx}        
\usepackage{multicol}        
\usepackage[bottom]{footmisc}
\usepackage[normalem]{ulem}	
\usepackage{hyperref}  
\usepackage{soul}   

\newcommand{\vx}{{\mathbf x}}

\newcommand{\dt}{{\Delta\tau}}
\def\v_op{ \hat{v} }
\def\vxbar{ \bar{\mathbf{x}} }

\newcommand\COMMENTED[1] {}

\makeindex             

\begin{document}

\title*{\emph{Ab Initio} Electronic Structure Calculations by Auxiliary-Field Quantum Monte Carlo}
\author{Shiwei Zhang}

\institute{Shiwei Zhang \at Center for Computational Quantum Physics, Flatiron Institute, New York, NY 10010 \\ 
\ AND \\ Department of Physics, College of William \& Mary, Williamsburg, VA 23185 \\
 \email{szhang@flatironinstitute.org}}



%
%
\maketitle
\abstract{The auxiliary-field quantum Monte Carlo (AFQMC) method provides a computational framework for solving the time-independent Schroedinger equation in atoms, molecules, solids, and a variety of model systems by stochastic sampling. We introduce the theory and formalism behind this framework, briefly discuss the key technical 
steps that turn it into an effective and practical computational method, present several illustrative results,
and conclude with comments on the prospects of \emph{ab initio} computation by this framework.}

\section{Introduction}

Predicting materials properties requires robust and reliable
calculations at the most fundamental level.  Often the effects
being studied or designed originate from electron             
correlations,
 and small errors in their treatment
can result in crucial and qualitative differences in the properties.            
The accurate treatment of interacting quantum systems
is one of the grand challenges in modern science. 
In condensed phase materials, the challenge is increased by the need to account for the interplay between the electrons and the chemical and structural environment. Progress in addressing this challenge will be fundamental to achieving the Òmaterials genomeÓ initiative.
 
Explicit solution of the many-body Schr\"odinger equation leads to
rapidly growing computational cost
as a function of system size (see, e.g., \cite{Szabo_book}).
To
circumvent the problem, most computational quantum mechanical studies
of large, realistic systems rely on simpler
independent-particle                                                
approaches based on density-functional theory (DFT) (see, e.g. \cite{DFT,Martin_book}),
using an approximate energy functional to include many-body
effects.                      
These replace the electron-electron interaction by an
effective potential, thereby reducing the problem to a set of
one-electron equations.          
Methods based on DFT and through  its
Car-Parrinello molecular dynamics implementation \citep{CarParMD}
have  been extremely effective                            
in complex molecules and solids \citep{DFT}.
Such approaches are the standard in
electronic structure, widely applied
in condensed matter, quantum chemistry, and materials science.

Despite the tremendous successes of  DFT,
the treatment of
electronic correlation                                            
is approximate.
For strongly correlated
systems (e.g., high-temperature superconductors,
heavy-fermion metals,                                          
magnetic materials, optical lattices),
where correlation effects
from particle interaction crucially modify materials properties,
the approximation can
lead to qualitatively incorrect results. Even in
moderately correlated systems
when the method is qualitatively correct, the
results are sometimes not sufficiently accurate.
For example, in ferroelectric materials
the usually acceptable 1\% errors
in DFT                                                                     
predictions of the
equilibrium lattice constant
can lead to almost full suppression of
the ferroelectric order.

The development of alternatives to independent-particle
theories is therefore of
paramount                                                           
fundamental and practical significance.
 To accurately capture
the quantum many-body effects, the size of the Hilbert space involved often
grows exponentially.
Simulation methods utilizing 
Monte Carlo (MC) sampling                                
\citep{Kalos1974,Foulkes01,CeperleyRMP95,BSS,Koonin,Zhang-Krak-2003}
 are, in principle, both
non-perturbative and well-equipped to handle details and complexities
in the external environment.                                                                                                                                                                                 
They are
a unique combination of accuracy, general applicability,
favorable  scaling (low-power) for computational cost with physical
system size, and scalability                                                                                                             
on parallel computing platforms  \citep{2008-SCIDAC-EndStation}.

For fermion systems, however, a
so-called ``sign'' problem \citep{schmidt84,Loh90,Zhang1999_Nato}
arises in varying forms in these MC simulation methods.
The
Pauli exclusion principle requires that the states be anti-symmetric
under interchange of two particles. As a consequence, negative signs
appear, which cause cancellations among contributions of the MC samples of the wave function or
density matrix. 
In some formalism, as discussed below, a phase appears which leads to a continuous degeneracy and
more severe cancellations.
As the temperature is lowered
or the system size is increased, 
such cancellation becomes more and
more complete. The net signal thus decays {\em exponentially\/} versus noise. The
algebraic scaling is then lost, and the method breaks down.  
Clearly the impact of this problem   
on the study of correlated electron systems is
extremely severe.

In this chapter, we discuss the auxiliary-field quantum Monte Carlo (AFQMC) method for many-body computations
in real materials. 
We cast the MC random walks in a 
space of over-complete Slater determinants, which significantly reduces the 
severity of the sign problem. In this space we formulate constraints on the random walk 
paths which lead to better approximations that are less sensitive to the details of the constraint.
We then develop internal checks and constraint release methods to systematically improve 
the approach. 
These methods have come
 under the name
of constrained path Monte Carlo (CPMC)           
\citep{Zhang1997_CPMC}
for systems where there is a sign problem (for
example, Hubbard-like models where the auxiliary-fields are real due to the short-ranged interactions ).
For electronic systems where there is a
phase problem (as the
Coloumb interaction leads to complex fields),
the methods 
have been referred to  as
phaseless AFQMC  \citep{Zhang-Krak-2003,Al-Saidi_GAFQMC_2006,WIRES-review}.
Here we will refer to the method as AFQMC; when necessary to emphasize 
the constrained-path (CP) approximation to distinguish the method from 
unconstrained free-projection, we will refer to it as CP-AFQMC.

\section{Formalism}

The Hamiltonian for any many-fermion system with two-body interactions
(e.g., the electronic Hamiltonian under the Born-Oppenheimer approximation) 
can be written as
\begin{equation}
   {\hat H} = {\hat H_1}+{\hat H_2} = 
  -\frac{\hbar^2}{2m} \sum_{m=1}^M \nabla_m^2
  + \sum_{m=1}^M V_{\rm ext}(\mathbf{r}_m)
+\sum_{m<n}^M V_{\rm int}(\mathbf{r}_m-\mathbf{r}_n)\,,
\label{eq:H_real}
\end{equation}
where ${\mathbf r}_m$ is the real-space coordinate of the $m$-th fermion.
The one-body part of the Hamiltonian, $ {\hat H_1}$, consists of the kinetic energy of the electrons
and the effect of the ionic (and any other external) potentials.
(We have represented the external potential as local, although this does not have to 
be the case. For example, in plane-wave calculations we will use a norm-conserving 
pseudopotential, which will lead to a non-local function $V_{\rm ext}$.) 
The two-body part of the Hamiltonian,  $ {\hat H_2}$, contains the electron-electron interaction terms.
The total number of fermions, $M$, will be fixed in the calculations we discuss. 
For simplicity, we have suppressed spin-index, but the spin will be made explicit when 
necessary. In that 
case, $M_\sigma$ is the
number of electrons with spin $\sigma$ ($\sigma=\uparrow$ or
$\downarrow$). We assume that the interaction is spin-independent, so the total 
$S_z$, defined by $(M_\uparrow - M_\downarrow)$, is fixed in the calculation, although it will be straightforward 
to generalize our discussions to treat other cases, for example, when there is 
spin-orbit coupling (SOC) \citep{Peter-SOC-review}. 

\index{Coloumb matrix elements}
With any chosen one-particle basis,
the Hamiltonian can be written  in second quantization 
 in the general form 
\begin{eqnarray}
{\hat H} = {\hat H_1}+{\hat H_2} 
= \sum_{i,j}^N {T_{ij} c_i^\dagger c_j}
   + {1 \over 2}
\sum_{i,j,k,l}^N {V_{ijkl} c_i^\dagger c_j^\dagger c_k c_l}\,,
\label{eq:H}
\end{eqnarray}
where 
the one-particle basis, $\{|\chi_i\rangle\}$ with $i=1,2,\cdots,N$, 
can be 
lattice sites
(Hubbard model),  plane-waves (as in solid state calculations) \citep{QMC-PW-Cherry:2007}, or Gaussians
(as in quantum chemistry) \citep{Al-Saidi_GAFQMC_2006,Wirawan_CaH2}, etc.
The operators
$c_i^\dagger$ and $c_i$ are creation and
annihilation operators on $|\chi_i\rangle$, satisfying standard fermion commutation relations.  The one-body matrix elements, $T_{ij}$,
contain the effect of both the kinetic energy and external potential, while the two-body
matrix elements, $V_{ijkl}$, are from the interaction. 
The matrix elements are 
expressed in terms of the basis functions,
for example, 
\begin{equation}
    V_{ijkl}
 = \int d\mathbf{r}_1 d\mathbf{r}_2
    \chi_i^*(\mathbf{r}_1) \chi_j^*(\mathbf{r}_2)
    V_{\rm int}(\mathbf{r}_1 -\mathbf{r}_2)   \chi_k(\mathbf{r}_2)                                                                              
       \chi_l(\mathbf{r}_1)\,.
\label{eq:Vmtx-def}
\end{equation}    
In quantum chemistry calculations, these are 
 readily evaluated with standard Gaussian basis sets.
In solid state calculations with plane-waves, the kinetic and electron-electron 
interaction terms have simple analytic expressions, while the electron-ion potential leads to 
terms which are provided by the pseudopotential generation.
We will assume that all matrix elements in Eq.~(\ref{eq:H}) have been evaluated and are known as 
we begin our many-body calculations.

\subsection{Non-orthogonal Slater determinant space}
\label{Sec_Slater}

The AFQMC method seeks to obtain the ground state of the Hamiltonian in Eq.~(\ref{eq:H}),
representing it stochastically in the form
\begin{equation}
    |\Psi_0\rangle=\sum_\phi \alpha_\phi |\phi\rangle\,,
\label{eq:GS-multi-det}
\end{equation}
where $|\phi\rangle$ is a 
Slater
determinant:
\begin{equation}
    |\phi\rangle \equiv {\hat \varphi}_1^\dagger 
{\hat \varphi}_2^\dagger \cdots
                   {\hat \varphi}_M^\dagger|0\rangle\,.
\label{eq:slater}
\end{equation}
In Eq.~(\ref{eq:slater}), the operator
$ {\hat \varphi}_m^\dagger \equiv \sum_i c_i^\dagger\, \varphi_{i,m}$, with $m$ taking an integer value
among $1,2, \cdots, M$,
creates an electron in a single-particle orbital $ \varphi_{m}$:
 $ {\hat \varphi}_m^\dagger  |0\rangle=\sum_i \varphi_{i,m} |\chi_i\rangle$.
 The content of the orbital can thus be  
 conveniently expressed as an $N$-dimensional vector
$\{ \varphi_{1,m}, \varphi_{2,m}, \cdots, \varphi_{N,m} \}$.
The Slater
determinant $|\phi\rangle$ in Eq.~(\ref{eq:slater}) 
can then be  expressed as an $N\times M$ matrix:
\begin{displaymath}
\Phi\equiv \left(\begin{array}{cccc}
\varphi_{1,1} & \varphi_{1,2} & \cdots & \varphi_{1,M}\\
\varphi_{2,1} & \varphi_{2,2} & \cdots & \varphi_{2,M}\\
      \vdots & \vdots & & \vdots\\
\varphi_{N,1} & \varphi_{N,2} & \cdots & \varphi_{N,M}
     \end{array}\right)\,.
\end{displaymath}
Each column of this
matrix represents a single-particle orbital that is completely
specified by its $N$-dimensional vector. 
For convenience, we will think of the  different columns as all properly 
orthonormalized, which is straightforward to achieve by, for example, 
modified Gram-Schmidt (see e.g., \cite{Zhang2000_BookChapter,Lecture-notes,WIRES-review}).

The mean-field Hartree-Fock (HF) 
solution is of course an example of a Slater determinant:
 $|\phi_{\rm HF}\rangle
=\prod_\sigma |\phi_{\rm HF}^\sigma\rangle$, where $|\phi_{\rm
HF}^\sigma\rangle$ is defined by a matrix $\Phi_{\rm HF}^\sigma$ whose
columns are the $M_\sigma$ lowest HF eigenstates. Similarly,
the occupied manifold in a DFT calculation forms a ``wave function'' which 
is a Slater determinant. 

\begin{figure}[b!]
 \includegraphics[width=1.\textwidth]{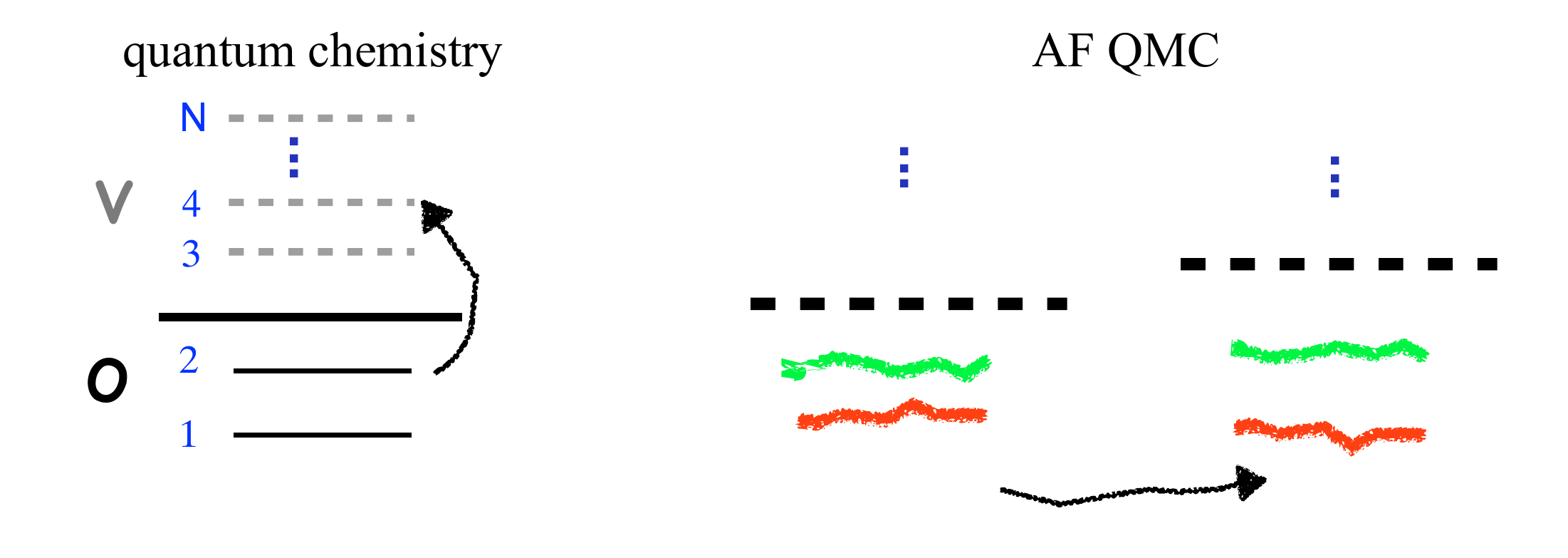}
 \caption{Schematic illustration of the connection and difference between quantum chemistry (QC)
 approaches and AFQMC. A fictitious system, with $M_\uparrow=2$, $M_\downarrow\le M_\uparrow$ and 
 $N$ basis functions, is shown. Vertical scale indicates single-particle energy.
 In QC-based methods (left), Slater determinants are constructed using
 the molecular orbitals from a (restricted-orbital) HF calculation. (``O" denotes occupied,
  ``V" denotes virtual, and the thick line indicates the Fermi level.) All resulting Slater determinants are orthogonal to each other, and to the 
 reference HF state. In AFQMC, each Slater determinant in Eq.~(\ref{eq:GS-multi-det}) 
 is sampled by random walk. Each walker $|\phi\rangle$ has only ``occupied'' orbitals (denoted by 
 the red and green lines) which are rotated during the random walk, under the influence of 
 stochastic auxiliary-fields (illustrated by the \emph{wiggly} lines, which move up and down as the walker evolves from one step to the next). The Slater determinants generated in AFQMC are non-orthogonal to each other.}
 \label{fig_QCvsAFQMC}
\end{figure}

In standard quantum chemistry (QC) methods, the many-body ground-state wave function is 
also represented by a sum of Slater determinants.
However, 
there is a key difference between it and the AFQMC representation. In QC methods, the different 
Slater determinants are orthogonal. As illustrated in the left panel in Fig.~\ref{fig_QCvsAFQMC},
each of the determinants is formed by
excitations from the HF determinant. In other words, each
 $|\phi\rangle$ on the right-hand side of Eq.~(\ref{eq:GS-multi-det}) 
 is given by a set of $M$ molecular orbitals (MOs), 
 and the corresponding matrix is formed by orthonormal \emph{unit vectors}.
 In contrast, in AFQMC the different Slater determinants on the right-hand side of Eq.~(\ref{eq:GS-multi-det}) are not orthogonal: $\langle \phi'|\phi\rangle\ne 0$. 
They are obtained by rotations of the occupied 
 orbitals using one-body Hamiltonians involving random auxiliary-fields (see further details below), as illustrated
 in the right panel in Fig.~\ref{fig_QCvsAFQMC}.  

Several properties of non-orthogonal Slater determinants are worth mentioning.
The overlap between two of them 
is given by
\begin{equation}
\langle \phi|\phi^\prime\rangle = \det(\Phi^\dagger\Phi^\prime).
\label{eq:ovlp}
\end{equation}
We can define the expectation of an operator ${\hat O}$
with respect to a pair of non-orthogonal Slater determinants: 
\begin{equation}
{\langle{\hat O}\rangle}_{\phi\phi'}
\equiv 
\frac{\langle\phi|{\hat O} |\phi^\prime\rangle}
                  {\langle\phi|\phi^\prime\rangle}\,,
\label{eq:expect_T=0}
\end{equation}
for instance single-particle Green's function
   $G_{ij} \equiv {\langle c_i c_j^\dagger\rangle}_{\phi\phi'}$:
\begin{equation}
   G_{ij} \equiv
\frac{\langle\phi|c_i c_j^\dagger |\phi^\prime\rangle}
                  {\langle\phi|\phi^\prime\rangle}
          = \delta_{ij}
          - [\Phi^\prime(\Phi^\dagger\Phi^\prime)^{-1}\Phi^\dagger]_{ij}.
\label{eq:G}
\end{equation}
Given the Green's function matrix $G$, 
the general expectation defined in Eq.~(\ref{eq:expect_T=0})
can be computed for most operators
of interest. 
For example, we can calculate the expectation  
of a general two-body operator, 
${\hat O}=\sum_{ijkl} O_{ijkl} c^\dagger_i c^\dagger_j c_k c_l$, 
under the definition of Eq.~(\ref{eq:expect_T=0}): 
\begin{equation}
{\langle{\hat O}\rangle}_{\phi\phi'} 
=\sum_{ijkl} O_{ijkl}
(G^\prime_{jk} G^\prime_{il} - G^\prime_{ik} G^\prime_{jl}),
\label{eq:expect}
\end{equation}
where the matrix $G^\prime$ is defined as $G^\prime\equiv I-G$. 

A key property of Slater determinants we will invoke is the \emph{Thouless Theorem}: 
 any one-particle 
operator ${\hat B}$ of the form
\begin{equation}
{\hat B}={\rm exp}\big(\sum_{ij}c_i^\dagger U_{ij}c_j\big)\,,
\label{eq:spo}
\end{equation}
when acted on a Slater determinant,
simply leads to another Slater determinant \citep{Hamann1990}, i.e.,
\begin{equation}
   {\hat B}|\phi\rangle =
         {\hat \phi}_1^{\prime\;\dagger} {\hat \phi}_2^{\prime\;\dagger} \cdots
         {\hat \phi}_M^{\prime\;\dagger}|0\rangle
         \equiv |\phi^\prime\rangle
\label{eq:expo}
\end{equation}
with ${\hat \phi}_m^{\prime\;\dagger} = \sum_j c_j^\dagger\,
\Phi^\prime_{jm}$ and $\Phi^\prime\equiv e^{U}\Phi$, where $U$
is a square matrix whose elements are given by $U_{ij}$ and
$B\equiv \exp(U)$ is therefore an $N\times N$ square matrix as well.
In other words, the operation of ${\hat B}$ on $|\phi\rangle$
simply involves multiplying an $N\times N$ matrix to the $N\times M$ matrix
representing the Slater determinant.

\index{single-particle Green's function}

There are several generalizations of the formalism we have discussed which extends the capability
and/or accuracy of the AFQMC framework. 
These can be thought of as generalizing 
one or both of the Slater determinants in Eqs.~(\ref{eq:ovlp}), 
(\ref{eq:expect_T=0}), and (\ref{eq:G}). From the viewpoint of AFQMC, as we shall discuss below, 
the``bra'' in these equations represents the trial wave function, and the ``ket'' represents 
the random walker:
\begin{itemize}

\item The first generalization is to replace $\langle \phi|$ by a projected Bardeen-Cooper-Schrieffer (BCS) wave function, that is, to use a projected BCS as a trial wave function, which can be advantageous 
for systems with pairing order. The corresponding overlap, Green
functions, and two-body mixed expectations have been worked out \citep{FG_pairing_AFQMC_PhysRevA.84.061602}. 

\item The second is to have both $\langle \phi|$ and $|\phi'\rangle$ in generalized 
HF (GHF) form, which is necessary to treat systems with spin-orbit coupling (SOC). The required modification 
to the formalism outlined above is given by \cite{Peter-SOC-review}. 

\item The third generalization is to have both sides in Hartree-Fock-Bogoliubov (HFB) form, for example, to
treat Hamiltonians with pairing fields. This will also be useful when using AFQMC as an impurity solver
in which the embedding  induces pairing order.
The corresponding AFQMC formalism has been described  \citep{Hao-HFB}.

\end{itemize}

\subsection{Ground-state projection}
\label{ssec:proj}

Most ground-state quantum MC (QMC) methods are based on iterative projection:
\begin{equation}
  |\Psi_0\rangle \propto \lim_{\tau\rightarrow\infty}
             e^{-\tau {\hat H}}|\Psi_T\rangle;
\end{equation}
that is, the ground state $|\Psi_0\rangle$ of a many-body 
Hamiltonian ${\hat H}$ can be projected from
any known trial state $|\Psi_T\rangle$ that satisfies $\langle
\Psi_T|\Psi_0\rangle\ne0$. 
In a numerical method, the limit can be obtained iteratively by
\begin{equation}
  |\Psi^{(n+1)}\rangle = e^{-\Delta\tau {\hat H}}|\Psi^{(n)}\rangle,
\label{eq:process}
\end{equation}
where $|\Psi^{(0)}\rangle = |\Psi_T\rangle$.  
Ground-state expectation $\langle {\hat
O}\rangle$ of a physical observable ${\hat O}$
is given by 
\begin{equation}
  \langle {\hat O} \rangle 
=\lim_{n\rightarrow\infty} 
{\langle \Psi^{(n)} |{\hat O}|\Psi^{(n)}\rangle \over
\langle \Psi^{(n)} | \Psi^{(n)}\rangle}.
\label{eq:expectO_T=0}
\end{equation}
For example, the ground-state energy can be obtained by 
letting ${\hat O}={\hat H}$. A so-called mixed estimator exists,
however, which is exact 
for the energy (or any other ${\hat O}$ that commutes with 
${\hat H}$) and can lead to 
considerable simplifications in practice:
\begin{equation}
 E_0
=\lim_{n\rightarrow\infty}
{\langle \Psi_T |{\hat H}|\Psi^{(n)}\rangle \over
\langle \Psi_T | \Psi^{(n)}\rangle}.
\label{eq:mixedE}
\end{equation}

QMC methods carry out the iteration in
Eq.~(\ref{eq:process}) by Monte Carlo (MC)
sampling. The difference between different classes of methods
amounts primarily to the space that is used to represent the wave
function or density matrix and to carry out the integration.  The
AFQMC methods
work in second quantized representation and in an
auxiliary-field space, while Green's function Monte Carlo (GFMC) or diffusion Monte Carlo (DMC) works in
first-quantized representation and in electron coordinate space \citep{Kalos1974,Foulkes01}. 
The full-configuration interaction QMC (FCIQMC) 
\citep{FCIQMC1_JCP09} works in orthogonal Slater determinant space as in QC methods.

\begin{figure}[b!]
 \centering
 \includegraphics[width=0.7\textwidth]{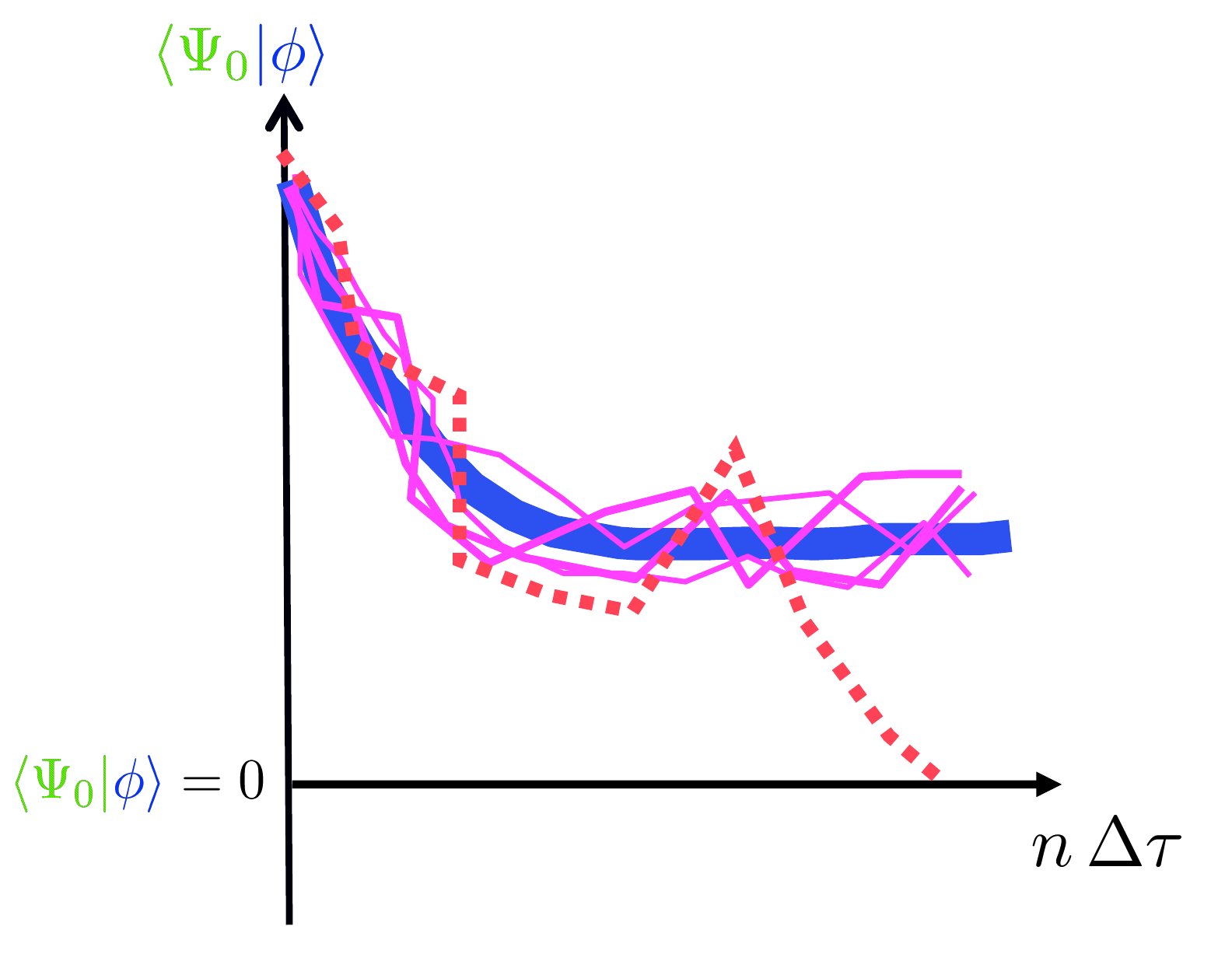}
 \caption{Illustration of the iterative imaginary-time projection to the ground state. 
 The overlap of the Slater determinants with a test wave function (e.g., the exact ground state $|\Psi_0\rangle$)
is plotted vs.~imaginary time $n\Delta\tau$.
 The thick blue line indicates a projection using $e^{-\Delta\tau\hat{H}_{\rm LDA}(\phi^{(n)})}$ 
 which converges to the LDA ground state (or a local minimum).                                                               
The wiggly magenta lines indicate an AFQMC projection which captures the many-body effect beyond LDA
as a stochastic linear superposition.
The propagator is obtained by expanding the two-body part of the $\hat{H}$, namely
$\hat{H}_2-\hat{V}_{xc}$, by a Hubbard-Stratonovich transformation as discussed in the next sections.
The dotted redline indicates a path which can lead to a sign problem (Sec.~\ref{SubSec:Phaseless}).
}
 \label{fig_proj-dft}
\end{figure}

Let us assume
that  $|\Psi_T\rangle$ is of the form of a single Slater determinant or a linear
combination of Slater determinants, as in Eq.~(\ref{eq:GS-multi-det}).
The operation of $e^{-\tau {\hat H_1}}$ on a Slater determinant 
simply yields another
determinant, per Thouless Theorem. The ground-state projection would therefore turn into
the propagation of a single Slater determinant if it were somehow possible to
write the two-body propagator $e^{-\tau {\hat H_2}}$
as the exponential of a one-body
operator. 

The above is realized in independent-electron theories. 
In the HF approximation, ${\hat H_2}$ is replaced by
one-body operators times expectations with respect to the current
Slater determinant wave function, schematically:
\begin{eqnarray}
c_i^\dagger c_j^\dagger c_k c_l \rightarrow c_i^\dagger c_l
\langle c_j^\dagger c_k\rangle
-c_i^\dagger c_k \langle c_j^\dagger c_l\rangle\,.
\end{eqnarray}
(A decomposition that includes pairing is also possible, leading to a Hartree-Fock-Bogoliubov
calculation.)
In the local density approximation (LDA) in 
DFT, 
${\hat H_2}$ is replaced by
$ \hat{H}_{\rm LDA}=
 {\hat H_1}+{\hat V}_{xc}$, where
 ${\hat V}_{xc}$ contains
  the density operator in real-space, with matrix elements given by the exchange-correlation 
functional which is computed with the local density from the current 
Slater determinant in the self-consistent process.
 In both these cases, an iterative
procedure can be used, following Eq.~(\ref{eq:process}), to project out the solution to the approximate
Hamiltonians, as an
imaginary-time evolution of a single Slater determinant \citep{Zhang2008_HFjell}.
This is illustrated by the blue line in Fig.~\ref{fig_proj-dft}.
Note that this procedure is formally very similar to time-dependent HF or time-dependent DFT (TDDFT), 
except for the distinction of imaginary versus real time.

\subsection{Hubbard-Stratonovich transformation}
\label{ssec:HS}

Suppose that ${\hat H_2}$, the two-body part in the Hamiltonian in Eq.~(\ref{eq:H}),
can be written as a sum of squares of one-body operators:
\begin{equation}
{\hat H_2}=\frac{1}{2}\sum_{\gamma=1}^{N_\gamma} \lambda_\gamma {\hat v_\gamma}^2, 
\label{eq:HS-H2_vsq}
\end{equation}
where $\lambda_\gamma$ is a constant, $\hat v_\gamma$ is a one-body operator similar to $\hat H_1$,
and $N_\gamma$ is an integer.
We can then apply the Hubbard-Stratonovich (HS) transformation to each term 
\begin{equation}
   e^{-\frac{\Delta\tau}{2}\:\lambda\, {\hat v}^2} 
= \int_{-\infty}^\infty dx
            \frac{e^{-\frac{1}{2} x^2}}{\sqrt{2\pi}}
           e^{x \sqrt{-\Delta\tau\lambda}\,{\hat v}},
\label{eq:HStrans}
\end{equation}
where $x$ is an auxiliary-field variable. 
The constant in front of $\hat v$ in
the exponent on the right-hand side can be real or imaginary, 
depending on the sign of $\lambda$. The key is that 
the quadratic form (in ${\hat v}$) on the left is replaced by a linear 
one on the right.
There are various ways to achieve the decomposition in Eq.~(\ref{eq:HS-H2_vsq})  for a general two-body 
term \citep{Negele_book}.
Below we outline the two most commonly applied 
cases in electronic structure: ({\bf a}) \emph{with planewave basis} and ({\bf b}) for a more dense matrix $V_{ijkl}$ resulting 
from a \emph{general basis set} such as Gaussians in QC.

In a \emph{plane-wave basis}, the two-body part is the Fourier transform of $1/|{\mathbf r}_m-{\mathbf r}_n|$
\citep{QMC-PW-Cherry:2007}:
\begin{equation}
\label{Vee}
\hat{H}_2 \rightarrow \frac{1}{{2\Omega }}
\sum\limits_{i,j,k,l} {\frac{{4\pi }}{{\left| {{\bf G}_{\bf i} {\bf - G}_{\bf k} } \right|^2
}}}  
 \ c_i^\dag c_j^\dag c_l c_k \delta _{{\bf G}_{\bf i} {\bf - G}_{\bf                                                        
k} {\bf ,G}_{\bf l} {\bf - G}_{\bf j} } \delta _{\sigma _i ,\sigma                                                                
_k } \delta _{\sigma _j ,\sigma _l }\,,
\end{equation}
where $\{ {\bf G}_{\bf i} \}$ are planewave wave-vectors, $\Omega$ is the volume of the supercell,
and $\sigma$ denotes spin.
Let us use ${\bf Q}\equiv {\bf G}_{\bf i} - {\bf G}_{\bf k}$, and define a density operator in momentum space:
\begin{equation}
\label{eq:rho_Q}
\hat{\rho}({\bf Q})\equiv 
 \sum\limits_{{\bf G},\sigma }
{c_{{\bf G + Q},\sigma }^\dag }
{c_{{\bf G},\sigma }^{ } } ,
\end{equation}
where the sum is over all ${\bf G}$ vectors which allow both ${\bf G}$  
and ${\bf G + Q}$ to  fall within the pre-defined kinetic energy 
cutoff in the planewave basis. The two-body term in Eq.~(\ref{Vee}) can then be manipulated into the form
\begin{equation}
\label{eq:PW-HS-ready}
\hat{H}_2 \rightarrow 
\sum\limits_{{\bf Q} \ne {\bf 0} } {\frac{{\pi }}{{\Omega Q^2 }}\,\left[\hat \rho^\dag({\bf Q})\,\hat \rho({\bf Q})
+\hat \rho({\bf Q})\,\hat \rho^\dag({\bf Q})\right]},
\end{equation}
where the sum is over all ${\bf Q}$'s except ${\bf Q}=0$, since in Eq.~(\ref{Vee})
the ${\bf G}_{\bf i} = {\bf G}_{\bf k}$ term is excluded
due to charge neutrality, and we have invoked $\rho^\dag({\bf Q})=\rho(-{\bf Q})$.
By making linear combinations of $ \left[(\rho^\dag({\bf Q})+\rho({\bf Q})\right]$
and $ \left[(\rho^\dag({\bf Q})-\rho({\bf Q})\right]$ terms,
we can then readily write 
the right-hand side in Eq.~(\ref{eq:PW-HS-ready}) in the desired square form of
Eq.~(\ref{eq:HS-H2_vsq}) \citep{QMC-PW-Cherry:2007}.

With a \emph{general basis} such as Gaussians 
yielding matrix elements given in Eq.~(\ref{eq:Vmtx-def}),
the most straightforward way to decompose $\hat H_2$ is through a direct diagonalization \citep{Al-Saidi_GAFQMC_2006,QMC-Hbonded_al-saidi:2007,Lecture-notes}. However, this is computationally costly. 
A modified Cholesky decomposition  
leads to ${\mathcal O}(N)$ fields \citep{Wirawan_CaH2,WIRES-review}. 
This approach, which has been commonly used in AFQMC for molecular systems with Gaussian basis sets and for downfolded Hamiltonians \citep{Ma-Downfolding-PRL}, proceeds as follows.
Let us cast $V_{ijkl}$ 
in the form of a two-index matrix
by introducing the compound indices
$\mu= (i, l)$ and $\nu=(j, k)$:  
$V_{\mu\nu}=V_{(i, l),(j, k)}= V_{ijkl}$. 
The  symmetric positive semidefinite matrix $V_{\mu\nu}$ is decomposed
using a recursive modified Cholesky algorithm                                                            
\citep{Koch2003,MOLCAS7}, to yield
\begin{equation}
\label{eq:modCD}
    V_{\mu\nu}                                                                     
= \sum_{\gamma= 1}^{N_\gamma}
    L_{\mu}^{\gamma}
    L_{\nu}^{\gamma}
  + \Delta_{\mu\nu}^{(N_\gamma)}\,,
\end{equation}
where $\Delta_{\mu\nu}^{(N_\gamma)}$ is the residual error at the $N_\gamma$-th iteration.
The iterative procedure is repeated until all matrix elements of the residual matrix are less than 
some pre-defined tolerance $\delta$:
\begin{equation}
\label{eq:modCD-tol}
    \left|V_{\mu\nu} - V_{\mu\nu}^{(N_\mathrm{CD})} \right|
=\left|                                                                   
        \Delta_{\mu\nu}^{(N_\mathrm{CD})}
    \right|
    \le
    \delta\,.
\end{equation}
For molecular calculations, typical values of $\delta$ range between $10^{-4}$ and $10^{-6}$ in atomic units
\citep{WIRES-review}.
Using the $N_\mathrm{CD}$ Cholesky vectors, we can re-write the two-body part of the Hamiltonian 
\begin{equation}
\label{V2-mCD}
\hat{H}_2 \rightarrow \frac{1}{{2}}
\sum\limits_{\gamma=1}^{N_\mathrm{CD}} 
 \left(
        \sum_{il} L_{\mu(i,l)}^\gamma c^\dagger_i c_l
    \right)
    \left(
        \sum_{jk} L_{\nu(j,k)}^\gamma c_j^\dagger c_k 
    \right) + {\cal O}(\delta)\,.
\end{equation}
Hence the form in Eq.~(\ref{eq:HS-H2_vsq}) is realized, with 
${\hat v_\gamma}=\sum_{il} L_{\mu(i,l)}^\gamma c^\dagger_i c_l$.

Different forms of the HS transformation  
 can affect the performance of the
AFQMC method. 
For example, it is useful to subtract a mean-field ``background'' from the two-body term prior 
to the decomposition \citep{Shiftcont1998,Purwanto2005,Al-Saidi_GAFQMC_2006}. 
The idea is that using the HS to decompose any constant shifts in the two-body interaction will 
necessarily result in more statistical noise. 
In fact, it has been shown \citep{Hao_symmetry_2012,WIRES-review} 
that the mean-field background subtraction 
can not only impact the statistical accuracy, but also lead to different
quality of approximations under the constrained path methods that we
discuss in the next section. 

If we denote the 
collection of auxiliary fields by $\vx$ and combine 
one-body terms from ${\hat H_1}$ and ${\hat H_2}$, 
we obtain the following compact representation of the outcome
of the HS transformation:
\begin{equation}
      e^{-\Delta\tau {\hat H}}= 
\int d\vx\,  p(\vx) {\hat B}(\vx),
\label{eq:HS}
\end{equation}
where $p(\vx)$ is a probability density function (PDF), for example, 
a multi-dimensional Gaussian. The propagator ${\hat B}(\vx)$
in Eq.~(\ref{eq:HS}) has a {\em special form\/}, namely, it
is a product of operators of the type in Eq.~(\ref{eq:spo}), with
$U_{ij}$ depending on the auxiliary field. The
matrix representation of ${\hat B}(\vx)$ will be denoted 
by $B(\vx)$. 

Note that the matrix elements of $B(\vx)$ can become complex, for example
when $\lambda$ in Eq.~(\ref{eq:HStrans}) is positive, which occurs in both of the forms discussed above.
Sometimes we will refer to this situation as having complex auxiliary fields, but it should be understood
that terms in the PDF of the HS transformation and ${\hat B}(\vx)$  can be re-arranged, and the 
relevant point is whether the Slater determinant has matrix elements which are real or complex, as further discussed in the next section.

In essence, the HS transformation replaces the two-body interaction by
one-body interactions with a set of random external auxiliary
fields. In other words, it converts an interacting system into many
{\it non-interacting} systems living in fluctuating external
auxiliary-fields.  The sum over all configurations of auxiliary fields
recovers the interaction.

\section{Ground-State AFQMC Methods}
\label{sec:AFQMC}

\subsection{Free-projection AFQMC} 
\label{sec:AFQMC_T=0}

We first briefly describe the ground-state 
AFQMC 
method without any constraints. 
Our goal is to illustrate the essential ideas, in a way
which will facilitate our ensuing discussions 
and help introduce the constrained path approximation and  the framework for the general AFQMC
methods that control the sign/phase problem.
We will not go into details, 
which are described in the literature.  

We write the usual path-integral and Metropolis form explicitly
here to show its connection to the open-ended random walk approach.
Ground-state expectation $\langle 
{\hat O}\rangle$ can 
by computed 
with Eqs.~(\ref{eq:process}) and (\ref{eq:HS}).
The denominator is
\begin{eqnarray}
  && \langle\psi^{(0)}| e^{-n\Delta\tau {\hat H}}\,
                     e^{-n\Delta\tau {\hat H}}|\psi^{(0)}\rangle
\nonumber\\
   &=& \int \langle \psi^{(0)}|
       \Bigl[\prod_{l=1}^{2n}
d\vx^{(l)} p(\vx^{(l)}){\hat B}(\vx^{(l)})\Bigr]|\psi^{(0)}\rangle
\nonumber\\
   &=& \int \Bigl[\prod_l d\vx^{(l)} p(\vx^{(l)})\Bigr]
         \det\Bigl([\Psi^{(0)}]^\dagger
\prod_l B(\vx^{(l)})\Psi^{(0)}\Bigr).
\label{eq:afqmc}
\end{eqnarray}
In the standard ground-state
AFQMC method \citep{Koonin,Sorella89,BSS}, a value of 
$n$ is first chosen and is kept fixed throughout the calculation.
If we use $X$ to denote the collection of the auxiliary-fields
$X=\{\vx^{(1)},\vx^{(2)},\dots,\vx^{(2n)} \}$ and $D(X)$ to represent the 
integrand in Eq.~(\ref{eq:afqmc}), we can write the
expectation value of Eq.~(\ref{eq:expectO_T=0}) as
\begin{equation}
\langle {\hat O} \rangle = 
{\int
{\langle {\hat O} \rangle}_{LR}\,D(X)\;dX
\over
\int D(X)\;dX}
=
{\int  
{\langle {\hat O} \rangle}_{LR}\,
\big| D(X)\big|\,\Theta(X)\;dX
\over
\int \big|D(X)\big|\,\Theta(X)\;dX},
\label{eq:AFQMC}
\end{equation}
where 
\begin{equation}
\Theta(X)\equiv D(X) /\big|D(X)\big|
\label{eq:sX}
\end{equation}
measures the phase of $D(X)$, which reduces to a sign when the overlap, $D(X)$,
is real along all paths $\{X\}$.
The  expectation for a given $X$, as  defined
in Eq.~(\ref{eq:expect_T=0}), is: 
\begin{equation}
{\langle {\hat O} \rangle}_{LR}\equiv
{\langle \phi_L| {\hat O} |\phi_R\rangle \over \langle \phi_L|\phi_R\rangle}
\label{eq:expect_1b_T=0}
\end{equation}
with
\begin{eqnarray*}
\langle\phi_L| & = &
\langle \psi^{(0)}|\,
{\hat B}(\vx^{(2n)}){\hat B}(\vx^{(2n-1)}) \cdots 
{\hat B}(\vx^{(n+1)})\\
|\phi_R\rangle & = & 
{\hat B}(\vx^{(n)}){\hat B}(\vx^{(n-1)}) \cdots {\hat B}(\vx^{(1)})
\,|\psi^{(0)}\rangle ,
\end{eqnarray*}
which are both Slater determinants.
 
$D(X)$ as well as $\langle\phi_L|$ and $|\phi_R\rangle$ are completely 
specified by the path $X$ in auxiliary-field space, given $|\Psi^{(0)}\rangle$.
The expectation in Eq.~(\ref{eq:AFQMC}) is thus a many-dimensional integration which can 
be evaluated by standard MC techniques.
Often the Metropolis algorithm \citep{Mal_book} is used to 
sample auxiliary-fields 
$X$ from $|D(X)|$.
We will refer to this as free-projection (in contrast with a constrained path
calculations). There are special Hamiltonians (e.g., repulsive Hubbard model at half-filling)
where special symmetry makes the sign problem absent. In those situations, the Metropolis 
approach described above is very effective and is the standard approach. It should be mentioned that in 
those cases, an infinite variance problem arises which must be controlled \citep{Hao-inf_var}.

\begin{figure}[b!]
 \centering
 \includegraphics[width=0.7\textwidth]{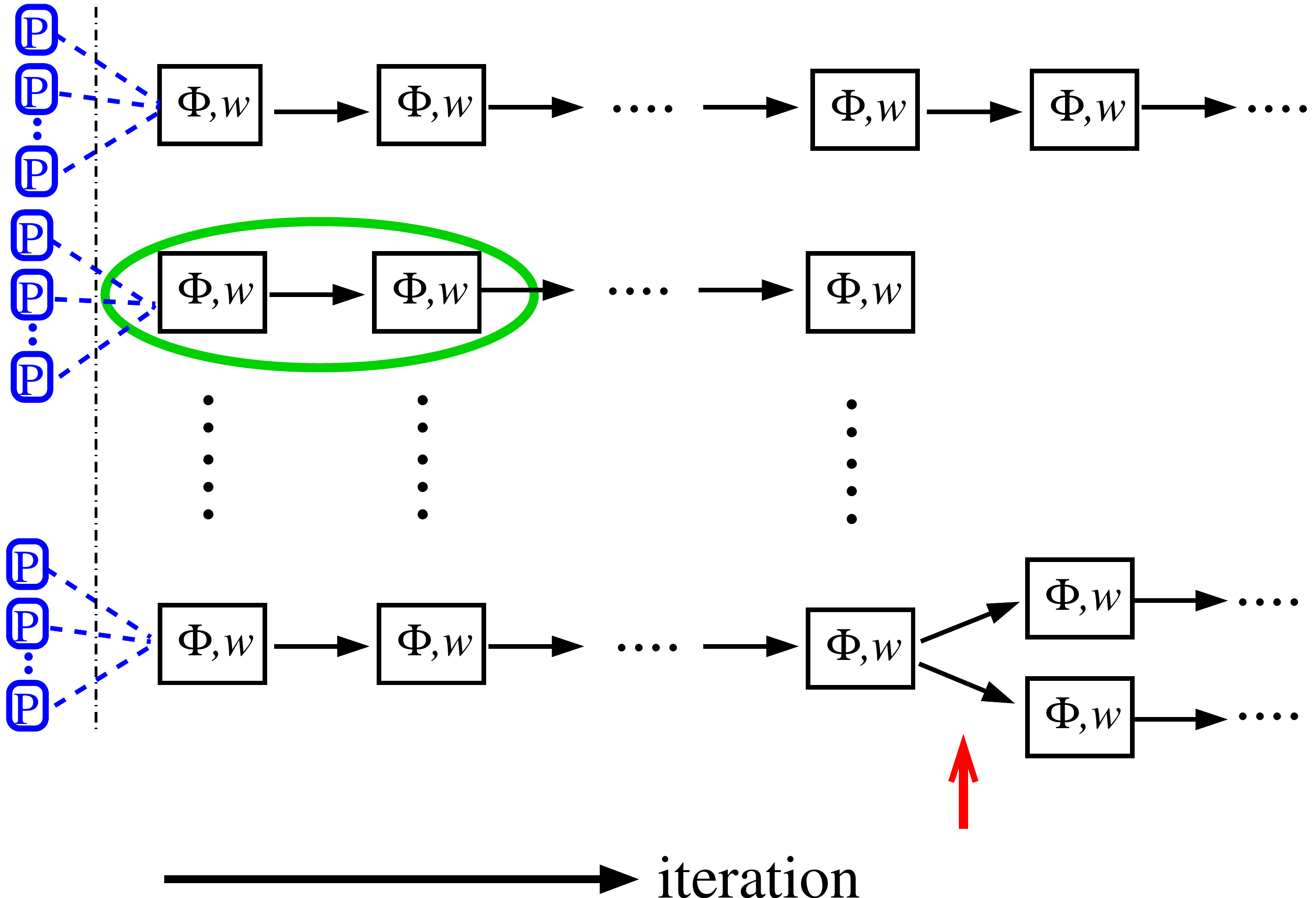}
 \caption{Schematic illustration of the AFQMC method.                                                              
Each box is a random walker, with
$|\phi\rangle$ the stochastic Slater determinant and $w$ its weight. A step (green oval)
is similar to one SCF (self-consistent-field) step in LDA. The red arrow indicates a population control,
where birth/death can occur. This structure allows for
exceptional capacity  for scaling on parallel computers. Multiple walkers can reside on one processor (blue 
``P'' box), or each walk can be split over
processors for large problems.}
 \label{fig_OpenRW}
\end{figure}

We carry out free-projection calculations with an open-ended random
walk \citep{Zhang1997_CPMC,Shiftcont1998} 
instead of using 
Metropolis sampling outlined above. 
For free projection calculations, the open-ended approach has no real advantage.
However, when a sign or phase problem is present, it is difficult to implement a
constraint to control the problem in the Metropolis framework, because of ergodicity 
issues  \citep{Fahy1990,Zhang1997_CPMC}.
The open-ended random walk framework avoids the difficulty, and is straightforward to project to longer imaginary-time
in order to approach the ground state. 
Moreover, when we carry out constraint release \citep{Hao_symmetry_2012}, 
the formalism will rely on the open-ended random walk. 
These points will become clear after we 
 illustrate the phase problem in electronic structure calculations below, and discuss
how the constraint can be formulated. The structure of the open-ended random walk 
is illustrated in Fig.~\ref{fig_OpenRW}.

\subsection{Constrained path AFQMC}
\label{SubSec:Phaseless}

As mentioned, a sign/phase problem occurs in the free-projection AFQMC, except for special cases
where the single-particle propagator $\hat B(\vx)$ satisfies particular symmetries 
(see, for example, \cite{PhysRevLett.116.250601}). In these cases, $\Theta(X)$ (defined in 
Eq.~(\ref{eq:sX})) vanishes, and 
$D(x)$ is real and non-negative.
Absent such special circumstances, a sign problem arises if  $\hat B(\vx)$ is real, and 
a phase problem arises if  $\hat B(\vx)$ is complex.
 As mentioned, the Coulomb interaction in $V_{\rm int}$ 
leads to a  phase problem in molecules and solids. 
In this section we discuss the constrained-path (CP) AFQMC, which for electronic systems 
has often been referred to as the phaseless or phase-free approximation  \citep{Zhang-Krak-2003,Purwanto2004}.

For real $\hat B(\vx)$ (e.g. Hubbard-type of short-range repulsive interactions decoupled with 
spin form of HS transformation), 
the sign problem occurs because of the fundamental symmetry between
the fermion ground-state $|\Psi_0\rangle$ and its negative
$-|\Psi_0\rangle$ \citep{Zhang1999_Nato,ZhangKalos_PRL}.  For any
ensemble of Slater determinants $\{|\phi\rangle\}$ which gives a MC 
representation of the ground-state wave function, as in Eq.~(\ref{eq:GS-multi-det}),
this symmetry
implies that
there exists another ensemble
$\{-|\phi\rangle\}$ which is also a correct representation.  In other
words, the Slater determinant space can be divided into two degenerate
halves ($+$ and $-$) whose bounding surface ${\cal N}$ is defined by
$\langle\Psi_0|\phi\rangle=0$. This dividing surface
is unknown. (In the cases with special symmetry mentioned above, the two sides separated 
by the surface are both positive. This has to do with the over-complete nature of the non-orthogonal 
Slater determinant space in AFQMC. A particular form of $\hat B(\vx)$ can pick out only a part of the 
space which can be non-negative.) 

\begin{figure}[t!]
 \centering
 \includegraphics[width=0.55\textwidth]{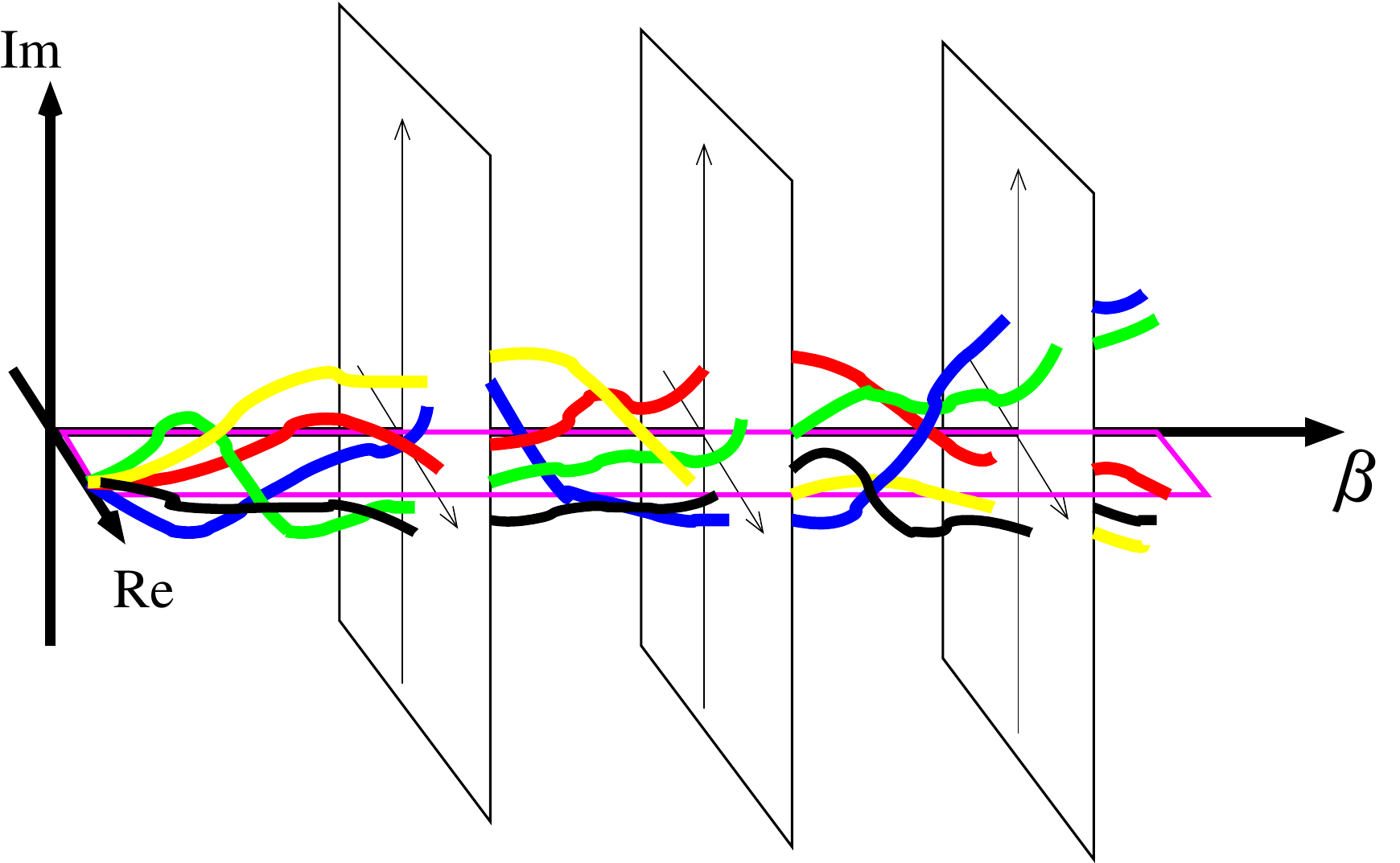}
\ \
\includegraphics[width=0.40\textwidth]{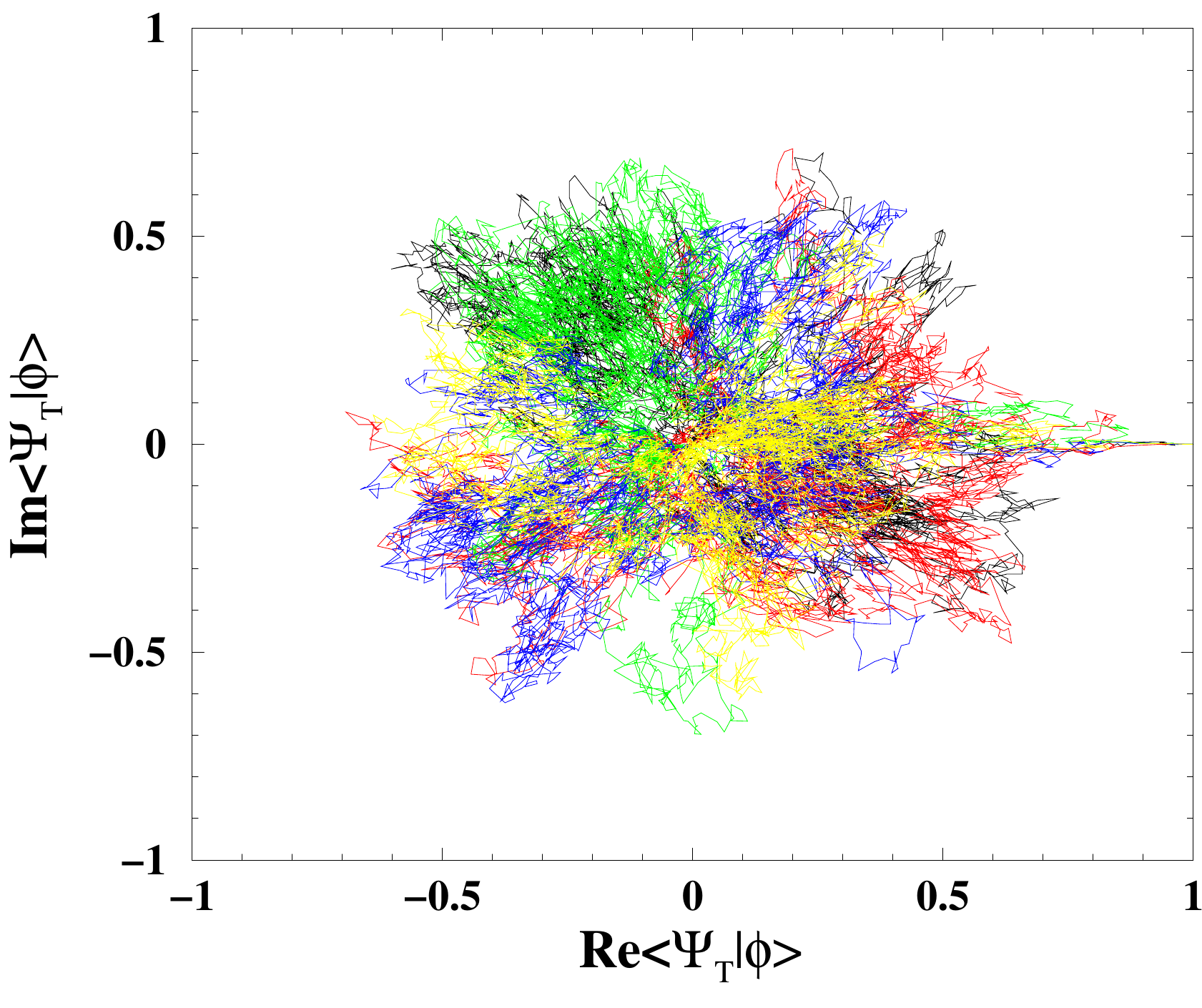}
 \caption{Schematic illustration of the phase problem and constraints to control
it.  The {\bf left panel}                                
shows, as a function of projection time $\beta\equiv n\dt$,
trajectories of 5 walkers (shown as 5 different colors) characterized by
the real (Re) and imaginary (Im) parts of their overlap with the
ground-state wave function.
The {\bf right panel}   
shows the walker distribution integrated over imaginary time,
i.e., the different frames
in the left panel stacked together along $\beta$.
The phase problem       
occurs because
the initial phase ``coherence'' of the random walkers
rapidly deteriorates with $\beta$, as they
become uniformly distributed
in the Re-Im-plane.
The idea of the phase constraint \citep{Zhang-Krak-2003} is to
apply a gauge transformation such that confining the random walk 
in the single  magenta plane (left)
is a
good approximation.}
  \label{figillusPhase}
\end{figure}

The idea of the $\pm$-symmetry can be seen from Fig.~\ref{fig_proj-dft}, where the 
dotted red line indicates a walker reaching the surface ${\cal N}$,
which will have a finite probability of occurring in a random walk, unless completely excluded by the dynamics. 
Once it does, it can, in general,
freely sample the two families of solutions which are symmetric about the horizontal axis
(above and below). A more detailed illustration and discussion of the sign problem can be found in 
 \cite{Zhang1999_FTCPMC} and \cite{Lecture-notes}.
 The idea of the phase problem is illustrated in Fig.~\ref{figillusPhase}. The complex plane 
 now replaces the vertical axis in Fig.~\ref{fig_proj-dft}, denoting the overlap of a random walker $|\phi\rangle$
 with the (hypothetically) known exact wave function, or the trial wave function $\langle \Psi_T|$.

Cancellation schemes can partially alleviate the problem, as demonstrated in 
coordinate space \citep{ZhangKalos_PRL,Anderson-cancel} and in Fock space \citep{FCIQMC1_JCP09}. 
To fully stabilize the calculation and restore polynomial scaling, however, an approximation has been necessary.
To date, the most effective and accurate method to achieve this has been the constrained path approach.

The method begins with a generalized similarity transformation in the spirit of the importance-sampling 
transformation. To make the derivation more concrete, we will use an explicit form of the 
HS transformation, and 
write the two-body propagator that results from Eq.~(\ref{eq:PW-HS-ready}) or (\ref{V2-mCD}) as
$\int e^{- {\vx^2 \over 2}} e^{\vx \cdot \hat{\mathbf v}} d\vx$,
where $\vx$ is an   $N_\gamma$-dimensional vector as given by Eq.~(\ref{eq:HS}) (with $\gamma$ labeling the auxiliary-fields) and
$\hat{\mathbf v}=\{\hat v_\gamma\}$ 
  denotes the collection of one-body operators. 
We introduce a shift 
to obtain an alternative propagator:
\begin{equation}
\int e^{- {\vx^2 \over 2} } e^{\vx \cdot \vxbar - {\vxbar^2 \over 2}}
e^{(\vx-\vxbar)\cdot \hat{\mathbf v}} d\vx\,,
\label{eq:prop_sig_shft}
\end{equation}
which is exact for any choice of the shift
$\vxbar$, including complex shifts.

We recall that the random walk is supposed to lead to a MC sampling of the coefficient $\alpha_\phi$
in Eq.~(\ref{eq:GS-multi-det}):
\begin{equation}
|\Psi_0\rangle \doteq \sum_{\{\phi\}} w_\phi |\phi\rangle\,.
\label{eq:wf_MC_free}
\end{equation}
The sum in Eq.~(\ref{eq:wf_MC_free}), which is over the population of walkers after an ``equilibration'' portion of the 
open-ended random walk has been discarded, is in a Monte Carlo sense, and is
typically much smaller than the sum in
Eq.~(\ref{eq:GS-multi-det}). The weight 
of each walker $|\phi\rangle$, $w_\phi$, can be thought of as $1$ (all 
walkers with equal weight); it is allowed to fluctuate only for practical (efficiency) consideration.

Using the idea of importance sampling, we seek to replace Eq.~(\ref{eq:wf_MC_free}) by the following 
to sample Eq.~(\ref{eq:GS-multi-det}):
\begin{equation}
|\Psi_0\rangle = \sum_\phi w_\phi
{|\phi\rangle \over \langle\Psi_T | \phi\rangle}\,,
\label{eq:wf_MC_imp}
\end{equation}
where any overall phase of the walker $|\phi\rangle$ is cancelled in the numerator and denominator 
on the right-hand side  \citep{Zhang-Krak-2003}.
This implies a modification to the propagator
in Eq.~(\ref{eq:prop_sig_shft}):
\begin{equation}
\int \langle \Psi_T|\phi^\prime(\vx)\rangle
e^{- {\vx^2 \over 2} } e^{\vx \vxbar - \vxbar^2/2}
e^{(\vx-\vxbar)\cdot \hat{\mathbf v}} {1 \over \langle\Psi_T | \phi\rangle}
d\vx,
\label{eq:prop_sig_imp}
\end{equation}
where 
$|\phi'({\mathbf x})\rangle=e^{({\mathbf x}-{\bar {\mathbf x}})\cdot \hat{\mathbf v}} |\phi\rangle$ and
the trial wave function $|\Psi_T\rangle$ represents the best
guess to 
$|\Psi_0\rangle$.
Let us define the following shorthand: 
\begin{equation}
\bar{\mathbf v} \equiv
- {\langle\Psi_T|\hat{\mathbf v}|\phi\rangle \over \langle\Psi_T | \phi\rangle}
\sim \mathcal{O}(\sqrt{\dt});\quad
\bar{{\mathbf v}^2} \equiv
{\langle\Psi_T|\hat{\mathbf v}^2|\phi\rangle \over \langle\Psi_T | \phi\rangle}
\sim \mathcal{O}(\dt)\,.
\end{equation}
We can then evaluate the ratio $\langle \Psi_T|\phi^\prime(\vx)\rangle            
/\langle\Psi_T | \phi\rangle$ in
Eq.~(\ref{eq:prop_sig_imp})
by expanding
the propagator \citep{Moskowitz1982,Zhang-Krak-2003,Purwanto2004} to $\mathcal{O}(\tau)$,
to obtain:
\begin{equation}
{\langle \Psi_T|\phi^\prime(\vx)\rangle
\over \langle\Psi_T | \phi\rangle}
e^{\vx \cdot \vxbar - \vxbar^2/2}
\doteq
\exp[- (\vx -\vxbar) \cdot \bar{\mathbf v}
+ {1 \over 2} (\vx -\vxbar)^2 \bar{{\mathbf v}^2}
- {1 \over 2} (\vx -\vxbar)^2 \bar{\mathbf v}^2 
+\vx \cdot \vxbar - \vxbar^2/2].
\label{eq:wt_raw}
\end{equation}

The optimal choice of the shift $\vxbar$, which we shall refer to as a force bias,
minimizes the fluctuation of Eq.~(\ref{eq:wt_raw}) with respect to
$\vx$, and it is straightforward to show that it is
$\vxbar=\bar{\mathbf v}$. 
With this choice, 
Eq.~(\ref{eq:prop_sig_imp})
can be written approximately as \citep{Lecture-notes}
\begin{equation}
\int e^{- {\vx^2 \over 2} }
e^{(\vx-{\bar{\mathbf v}})\cdot \hat{\mathbf v}} e^{ \bar{{\mathbf v}^2} \over 2}
d\vx.
\label{eq:prop_sig_El0}
\end{equation}
Restoring $\hat{H_1}$, we obtain 
the complete propagator:
\begin{equation}
\int e^{- {{\vx}^2 \over 2} }\,
\exp[-{\dt \hat{H_1} \over 2}]
\,\exp[({\vx}-{\bar{\mathbf v}}) \cdot \hat{\mathbf v}]\,\exp[-{\dt \hat{H_1} \over 2}]
\:\exp[-\dt E_L(\phi)]
\:d\vx,
\label{eq:prop_sig_El}
\end{equation}
where $E_L$ is the local energy, the mixed-estimate of the Hamiltonian:
\begin{equation}
E_L(\phi)\equiv
{\langle\Psi_T|\hat{H}|\phi\rangle \over \langle\Psi_T | \phi\rangle}.
\label{eq:El}
\end{equation}

In the limit of an exact
$|\Psi_T\rangle$, $E_L$ is a {\em real\/} constant, and the weight of
each walker remains real.  The mixed estimate for the
energy from Eq.~(\ref{eq:mixedE})
 is phaseless:
\begin{equation}
E_0^{\rm c} = {\sum_\phi w_\phi E_L(\phi) \over \sum_\phi w_\phi}.
\label{eq:mixed_w_EL}
\end{equation}
With a general $|\Psi_T\rangle$ which is not exact, a natural
approximation is to replace $E_L$ in Eq.~(\ref{eq:prop_sig_El}) by its
real part, ${\rm Re} E_L$. The same replacement is then necessary in
Eq.~(\ref{eq:mixed_w_EL}). 

When $\hat B(\vx)$ (i.e., $\hat{\mathbf v}$)
is real, this formalism
reduces to the so-called constrained-path approximation 
 \citep{Zhang1997_CPMC}.
Regardless of whether $\hat{\mathbf v}$ is real, the shift
$\vxbar$ diverges as the random walk in the complex
plane (see the  right panel of
Fig.~\ref{figillusPhase}) approaches the origin,  i.e., as $\langle\Psi_T|\phi^\prime\rangle                               
\rightarrow 0$.
 The effect of the divergence is to move the walker
away from the origin. With a {\em real\/} $\hat{\mathbf v}$, 
the random walkers move only on the real axis. If they are
initialized to have positive overlaps with $|\Psi_T\rangle$,
$\vxbar$ will ensure that the overlaps remain positive throughout
the random walk.

For a general case with a complex $\hat{\mathbf v}$, however, the
 formalism above  by itself is not sufficient to remove the phase problem. 
To see this we consider 
the phase of $\langle\Psi_T                                              
|\phi^\prime(\vx-\vxbar)\rangle/\langle\Psi_T |\phi\rangle$, which we denote by
$\Delta\theta$. In general, $ \Delta\theta \sim \mathcal{O}(-\vx {\rm           
Im}(\vxbar)$) is non-zero. This means that the walkers will undergo a random
walk in the complex plane. 
At large $\beta$ they will
therefore populate the complex plane symmetrically, independent of
their initial positions. 
This is illustrated in the  right panel of
Fig.~\ref{figillusPhase}, which shows $\langle                                      
\Psi_T|\phi\rangle$ for a three-dimensional jellium model with two electrons at $r_s=10$ for a total
projection time of $\beta=250$ (taken from \cite{Lecture-notes}). 
The random walk is ``rotationally invariant'' in the
complex plane, resulting in
 a vanishing signal-to-noise ratio asymptotically, even though
the walkers are all real initially with  
$\langle                                                                    
\Psi_T|\phi^{(0)}\rangle=1$.
An alternative but related way to state the problem is that,
despite the divergence of $\vxbar$,  
the build-up of a finite density at the origin of the complex plane cannot be prevented, unlike 
in the one-dimensional situation (real $\langle \Psi_0|\phi\rangle$, sign problem).  Near the
origin the local energy $E_L$ diverges, which causes diverging
fluctuations in the weights of walkers when the density does not vanish.

Thus the second ingredient of the constraint for the phase problem is to project 
the random walk back to
``one-dimension.'' This is done by reducing the weight of the walker 
in each step by the angular deviation of the overlap in the complex plane:
\begin{equation}
w_{\phi^\prime} \rightarrow w_{\phi^\prime} \max\{0,\cos(\Delta\theta)\}\,.
\label{eq:cos-proj}
\end{equation}
A prerequisite for this approximation to work well is the importance sampling transformation,
which has eliminated the leading order in the overall phase of $|\phi\rangle$ in the 
propagator in Eq.~(\ref{eq:wt_raw}). Given the transformation, several alternative forms
to the projection in Eq.~({\ref{eq:cos-proj}) were found to give similar accuracy  \citep{Zhang-Krak-2003,AFQMC-CPC2005,Purwanto2005,Lecture-notes}.

We can now summarize each step in the constrained path AFQMC formalism as
follows. For each random walker $|\phi\rangle$ in the current population $\{ |\phi\rangle, w_\phi\}$,

\newcounter{counter1}
\begin{list}{{\bf (\alph{counter1})}}{\usecounter{counter1}}

\item sample $\vx$ and propagate the walker
to $|\phi^\prime\rangle$
\begin{equation}
|\phi\rangle \rightarrow |\phi^\prime\rangle =
\exp[-{\dt \hat{H_1} \over 2}]
\,\exp[({\vx}-{\bar{\mathbf v}}) \cdot \hat{\mathbf v}]\,\exp[-{\dt \hat{H_1} \over 2}]
\:|\phi\rangle,
\label{eq:step_a}
\end{equation}

\item
update the weight of the walker
\begin{equation}
w_\phi \rightarrow w_{\phi^\prime} =
w_\phi \exp\Big[-\dt\cdot{\rm Re}\Big(E_L(\phi^\prime) + E_L(\phi)\Big)/2
\Big]\cdot\max\{0,\cos(\Delta\theta)\}\,.
\label{eq:step_b}
\end{equation}

\end{list}
Walkers so generated represent the
ground-state wave function with importance sampling, in the sense of
Eq.~(\ref{eq:wf_MC_imp}).

For additional technical details, we refer the reader to \cite{Lecture-notes,WIRES-review}, and references therein, for examples re-orthogonalization procedures  \citep{White1989,Zhang2000_BookChapter} 
to stabilize the walkers against numerical roundoff errors 
during the propagation, population control \citep{Umrigar1993,Zhang1997_CPMC} to regularize
the branching process, a hybrid alternative \citep{silicon_betatin_Purwanto2009}
to the local energy formalism in Eq.~(\ref{eq:prop_sig_El0}) 
to reduce computational cost in evaluating $E_l$, correlated sampling \citep{Shee-JCTC-2017}, 
constraint release  \citep{Hao_symmetry_2012} etc.

\subsection{Back-propagation for observables and correlation functions}
\label{SubSec:BP}

To calculate a correlation function or the expectation value of an observable which does not commute with
 the Hamiltonian, the mixed estimate in Eq.(\ref{eq:mixedE}) is biased, and the full estimator in 
 Eq.~(\ref{eq:expectO_T=0}) needs to be computed. 
 In the path-integral form in Eq.~(\ref{eq:AFQMC}), this is straightforward. With the open-ended 
 random walks, however, it is slightly more involved. A
  back-propogation (BP) technique \citep{Zhang1997_CPMC,Purwanto2004,Mario-BP} is employed. 

The idea of the BP  is to create two coupled populations to represent the bra and ket 
in Eq.~(\ref{eq:expectO_T=0}), respectively. Because the population in the random walk 
is importance-sampled, two independent populations which are uncoupled would 
lead to large fluctuations in the estimator after  the importance functions have been "undone"
 \citep{Purwanto2004}.
In BP, we choose an iteration $n$
and store the entire population
$\{\,|\phi^{(n)}_k\rangle,w^{(n)}_k\,\}$, where $k$ labels the walker in the population.  As the random walk proceeds from $n$, we keep
track of the following two items for each new walker: (1) the sampled
auxiliary-field values that led to the new
walker from its parent walker and (2) an integer label that identifies the
parent. After an additional $m$ iterations, we carry out the
back-propagation: For each walker $l$ in the $(n+m)$-th                   
(current) population, we initiate a determinant $\langle \psi_T|$ and act
on it with the corresponding propagators, but taken in reverse
order. The $m$ successive propagators are constructed from the 
items stored between steps $n+m$ and $n$, with ${\rm exp}(-\Delta\tau \hat{H_1}/2)$ inserted where necessary. The
resulting determinants $\langle \bar\phi^{(m)}_l|$ are combined with
its parent from iteration $n$, $|\phi^{(n)}_k\rangle$, to compute $\langle{\cal O}\rangle_{\rm           
BP}$, where $k$ is the index of the walker at step $n$ from which walker $l$ at step $(n+m)$ descended.  

In molecular systems, an improvement over the standard procedure has been proposed \citep{Mario-BP}.
The approach, called back-propagation with path restoration (BP-PRes),  allows one to ``undo'' some of the effect of the constraint in the forward direction (the $\cos$ projection
and the omission of the phase in $E_l$ in the weight). This reduced the effect of the constraint, which is applied in the forward 
direction  and does not preserve reversal
symmetry in imaginary time. 
With these advances, accurate observables and atomic forces have been obtained in molecules, 
paving the way for geometry optimization and \emph{ab inito} molecular dynamics with AFQMC.

Another recent development in methodology is the computation of imaginary-time correlation functions and
excitations. The techniques \citep{ettore-gap-Hubbard},
which have been applied in model systems and 
ultracold atoms so far, 
are directly applicable to real materials.

\section{Illustrative Results}

\begin{figure}[t!]
\begin{center}
\includegraphics[width=0.6\textwidth]{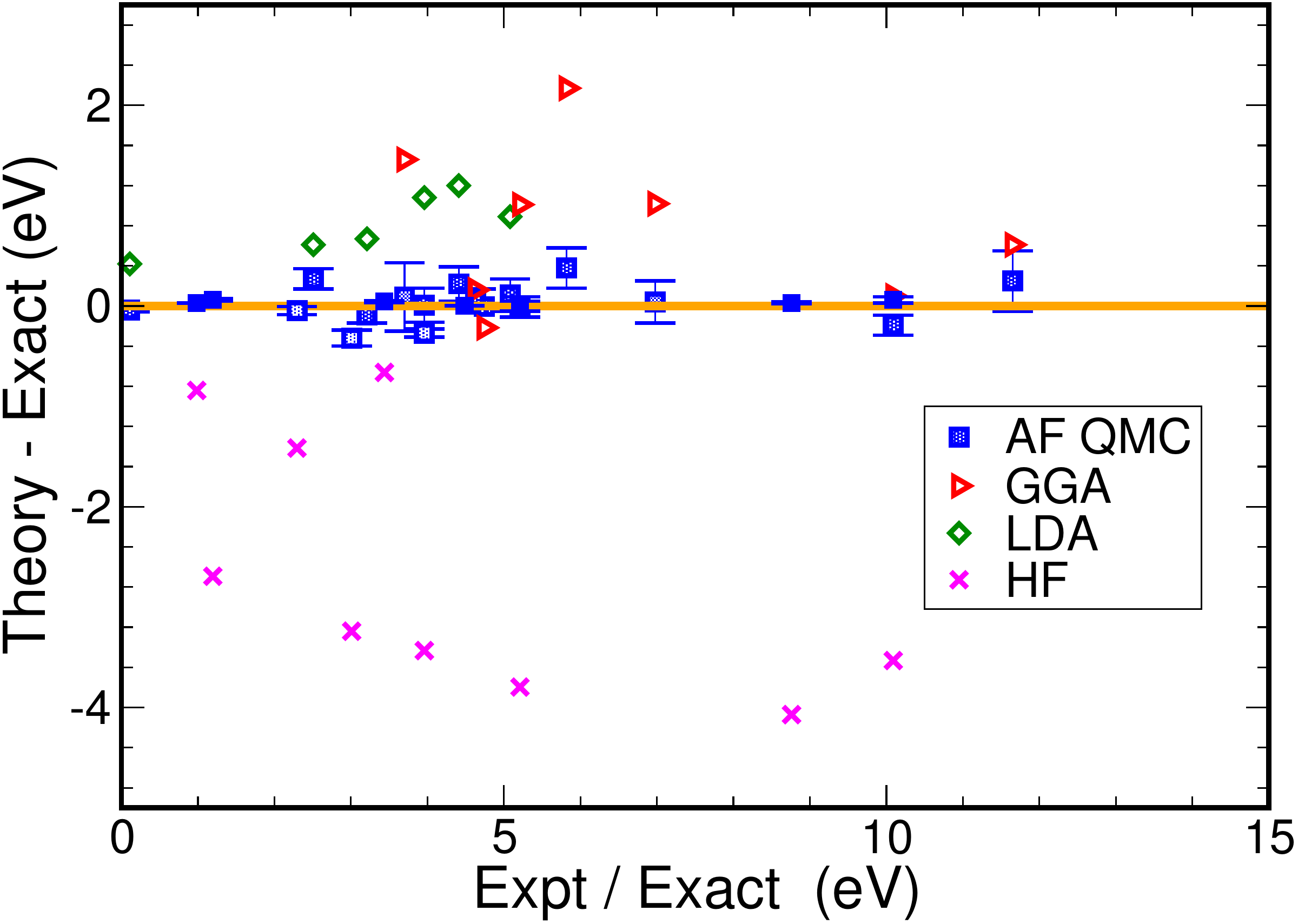}
\end{center}
\caption{
Calculated binding energies 
of molecules compared with
  experimental values  (taken from Refs~\cite{Lecture-notes,2008-SCIDAC-EndStation}).
  The discrepancy between theory and experiment
  is plotted. 
  Included are
 $sp$-bonded molecules, first- and second-row post-$d$ elements, and transition metal oxides.
 Several different forms of the AFQMC calculations was tested, including
 all-electron Gaussian basis sets, Gaussian basis with effective-core potentials, and plane-wave with
 pseudopotential.
   The AFQMC is fed a trial wave
  function to start, which is taken directly from DFT [with either
  LDA or the generalized-gradient
  approximation (GGA) functionals] or HF. The
  corresponding DFT or HF results are also shown.  As can be readily
  observed, the AFQMC results are in excellent agreement with
  experiment and significantly improve upon the values from DFT
  and HF.
}
\label{fig:Mol_binding}
\end{figure}

The AFQMC method has been applied to lattice models,  realistic solids
(using plane-wave basis and pseudo potentials),  molecular systems 
(using Gaussian basis sets), and  down-folded model Hamiltonians of real materials
(using DFT orbitals as basis sets). 
The method is just coming into form, and rapid advances in algorithmic development and in applications are
on-going. 
We briefly mention a few 
examples here to provide a flavor of how it has  been applied to date to tackle problems of 
electron correlations in materals.

For lattice models, most of the applications involve 
``only'' a sign problem, because of the short-range nature of the interaction. Here the 
constraint has no $\theta$ projection and reduces to a sign constraint. A large body of 
results exist, including recent benchmark results \citep{Hubbard-benchmark}.
Systems of ${\cal O}(1000)$ electrons have been treated quite routinely.
The AFQMC method has demonstrated 
excellent capabilities and accuracy, illustrating its potential as a general  
many-body computational paradigm. 
A key recent development \citep{mingpu-sc-cp} is to use the density or 
density matrix computed from AFQMC as a feedback into a mean-field calculation. 
The trial wave function $|\Psi_T\rangle$ obtained from the mean-field is then fed back into the 
AFQMC as a constraint, and a self-consistent constraining condition is achieved. This has lead
to further improvement in the accuracy and robustness of the calculation \citep{mingpu-sc-cp,science-stripe}.

For molecular systems, a recent review article is available \citep{WIRES-review} which 
describes in more detail the application of  AFQMC in quantum chemistry.
The formulation of AFQMC with Gaussian basis sets 
 has been extremely valuable.  Direct comparisons can be made with high-level QC results,
 which have provided valuable benchmark information and have been 
 crucial in gauging the AFQMC method as
 a general approach.
Figure~\ref{fig:Mol_binding} illustrates the results on molecules
using both planewave plus pseudopotentials and Gaussian basis sets.
In these calculations we have operated largely in an automated mode,
inputting only the DFT or HF 
solutions as $|\Psi_T\rangle$. This illustrates a potential mode of operation for AFQMC as a ``post-processing'' approach for molecules and
solids where additional accuracy is desired beyond standard DFT.

A benchmark study \citep{H-bbenchmark-PRX-2017}
was recently carried out involving 
a large set of modern many-body methods. 
AFQMC was among the methods included; consistent with previous findings, 
 the accuracy of 
AFQMC is found to be comparable to  CCSD(T),
the gold standard in chemistry \citep{bartlett-RMP2007,CC_Review_Crawford:2000},
near equilibrium geometry. For bond breaking, AFQMC was able to maintain 
systematic accuracy. Large basis sets and system sizes were reached and  an accurate equation of state was obtained.

\begin{figure}[t!]
  \begin{minipage}[t]{3.0in}
  \vskip0.15in
\includegraphics[width=0.8\textwidth]{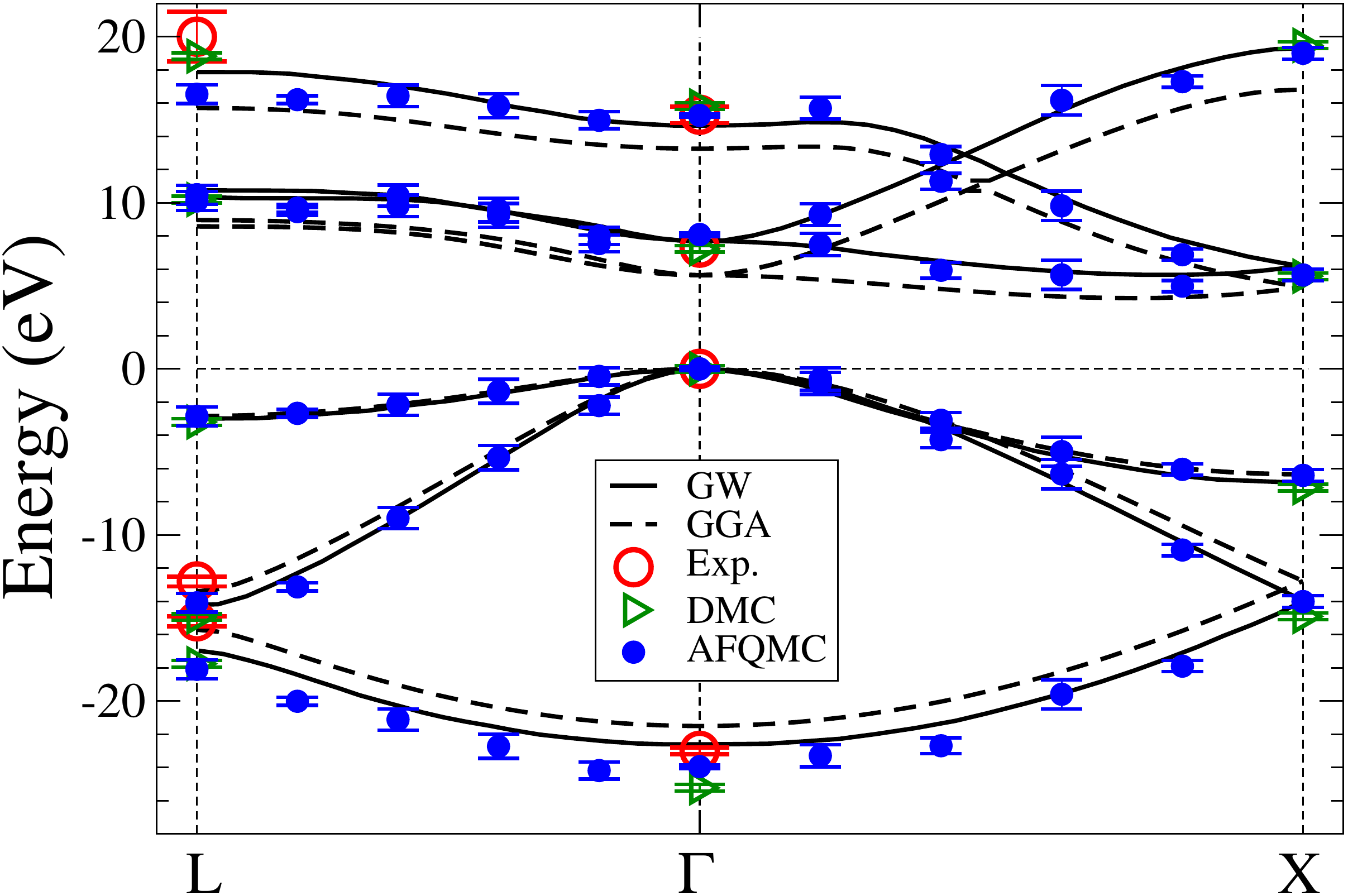}      
  \end{minipage}                                                                                      
  \begin{minipage}[t]{0.9in}                                                                                     
 \vskip0.15in
    \begin{tabular}{ll}
\hline\hline
  & $\Delta_{\rm gap}$ (eV) \\
\hline
GGA & 0.77 \\
Hybrid & 3.32,\,2.90 \\
LDA$+U$ & 1.0\\
$GW$ &  3.4,\,3.6,\,2.56 \\                                                    
AFQMC & 3.26(16)\\ 
expt. & 3.30,\,3.44,\,3.57\\
\hline\hline
\end{tabular}
\end{minipage}                                                           
\caption{
Computation of excitations and many-body quasi-particle band structures (taken from \cite{Ma_Ex_Solids_2012}).
The figure presents results on the band gap in diamond. Blue is AFQMC results,
$GW$ and DFT
band structures are plotted by solid and dashed lines, respectively.
Diffusion Monte Carlo (DMC)  results at
high symmetry points $\Gamma$, $X$, and $L$  are indicated by green triangles.
Experimental values  are shown as red circles.
The table shows the calculated fundamental band gap of wurtzite ZnO, compared with
experiment (and three DFT-based methods and $GW$).
}
\label{fig:solids_band}
\end{figure}

\begin{figure}[t!]
\includegraphics[scale=.47]{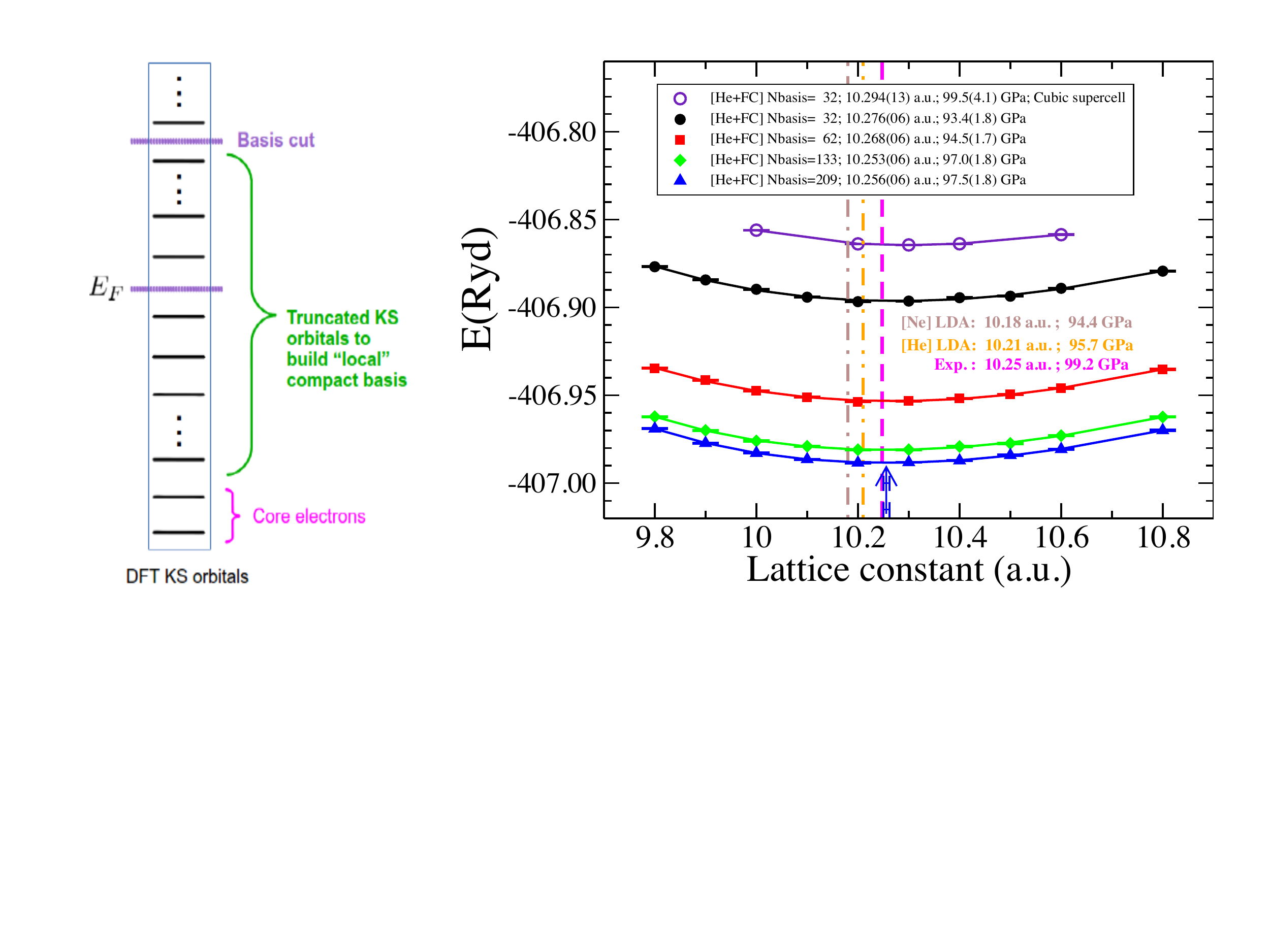}
\vskip-1.1in
\caption{
Pseudopential-free calculations in solids, and
a simple ``down-folding" approach to generate realistic model hamiltonians.
The {\bf left panel} illustrates a scheme to use Kohn-Sham (KS) orbitals obtained from a DFT
calculation as basis set for AFQMC. The DFT is performed with a plane-wave basis using
a helium-core. Keeping all the KS orbitals, including virtual orbitals,
below a certain cutoff ("Basis cut"), we compute the matrix elements with these orbitals as basis,
to obtain a many-body hamiltonian, which is then fed into the AFQMC. In the AFQMC, the
KS orbitals corresponding to the neon-core are frozen \citep{Wirawan-FC-JCTC13}.
In the {\bf right panel}, the calculated  equation of state  \citep{Ma-Downfolding-PRL}                               
is shown for a sequence of "Basis cut"
values.  
The calculated equilibrium lattice constant and bulk modulus are in excellent agreement with
experiment.}
\label{fig:downfold}      
\end{figure}

The AFQMC method can be used to study excited states.
Excited states distinguished by different symmetry from the ground state can be computed in a manner
similar to the ground state. For other excited states,
prevention of collapse into the ground state and control of the
fermion sign/phase problem are accomplished by a
 constraint using an excited state trial wave function \citep{2009-C2-exc-afqmc}. An additional orthogonalization constraint is formulated to use virtual orbitals  in solids for band structure calculations \citep{Ma_Ex_Solids_2012}. These constraints are not as ``clean'' or rigorous as that for the ground state.
 Use of improved  trial wave functions (for example, multi-determinant $|\Psi_T\rangle$ in molecules) 
 and the imposition of symmetry properties \citep{Hao_symmetry_2012} often lead to  improved results. 
 Tests in the challenging case of the
C$_2$ molecule yielded  spectroscopic constants
 in excellent agreement with experiment \citep{2009-C2-exc-afqmc}. 
 In Fig.~\ref{fig:solids_band} results from an
 application in solids are shown for the diamond band structure and for the 
 fundamental band gap in wurtzite ZnO \citep{Ma_Ex_Solids_2012}.

Planewave calculations in AFQMC can be built on standard 
plane-wave technologies in DFT calculations, as we have outlined. Norm-conserving pseudopotentials, 
including multiple-projector oseudopotentials, can be implemented straightforwardly \citep{Multiple-proj-Ma}. 
In order to reduce the cost of full planewave AFQMC calculations, a downfolding approach \citep{Ma-Downfolding-PRL} 
has been developed. The idea is to use Kohn-Sham orbitals (occupied and virtual) as basis sets.
The approach is illustrated in Fig.~\ref{fig:downfold}. 
The size of  the basis set in the largest calculation on the right, after downfolding, is more than an order of magnitude smaller than the total number of plane-waves, leading to large savings in the AFQMC computation.

\begin{figure}[t!]
\includegraphics[scale=.47]{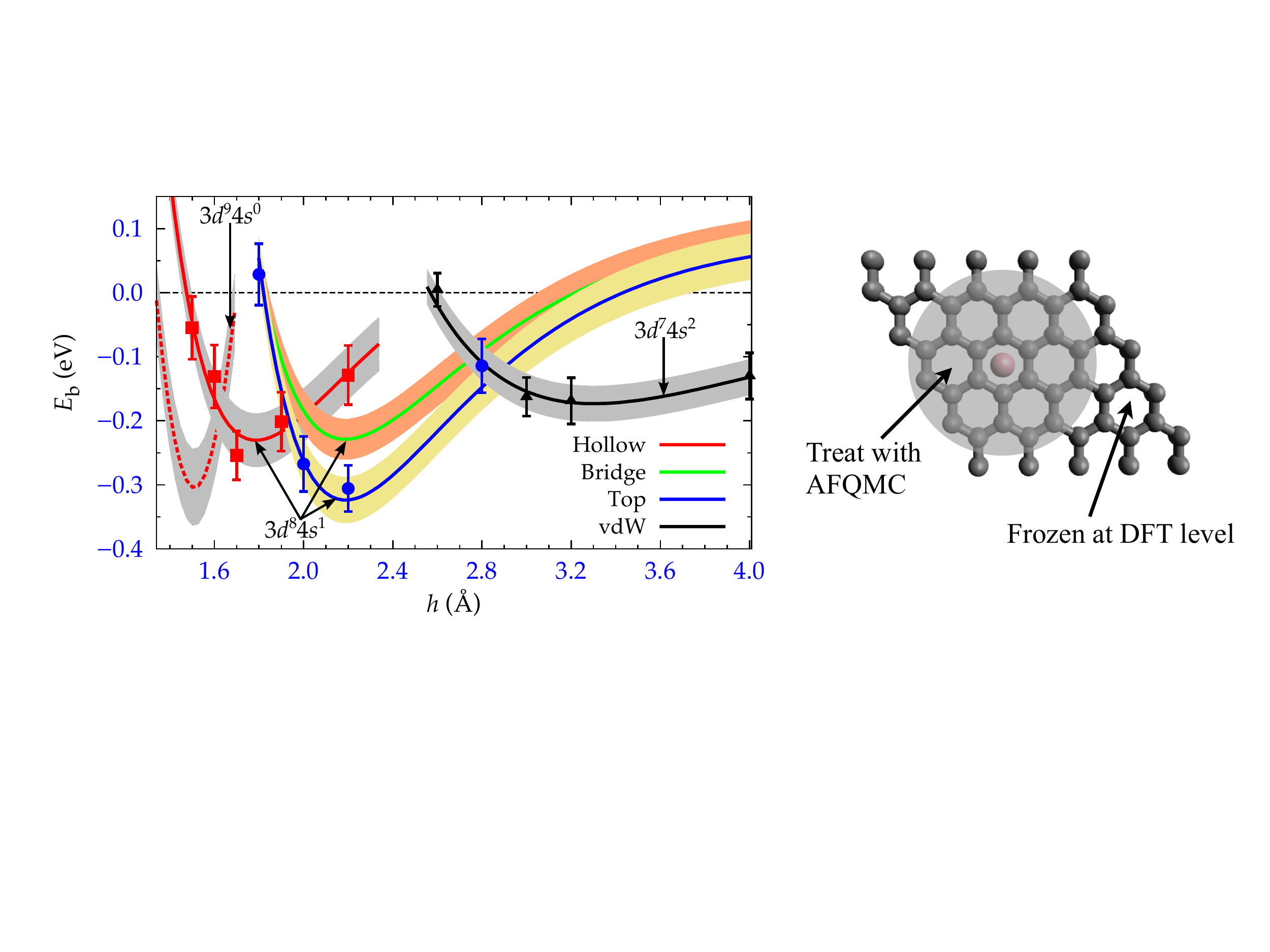}
\vskip-1.1in
\caption{
Co adsorption on graphene and embedding AFQMC within DFT.
The figure (taken from \cite{Co-graphene-Yudis-PRL}) shows
computed binding energy of Co on graphene, as a function of the distance $h$ between the
Co atom and the graphene plane. 
Squares, diamonds, circles, and triangles correspond to hollow (H, as illustrated on the right), 
bridge (B), and top (T) sites, and the van der Waals region, respectively. 
The dashed line indicates the low-spin H site (open squares).
Shaded areas are one-$\sigma$ estimates of uncertainties, including the statistical errors in AFQMC.
The right illustrates the embedding scheme for these calculations. The ``inner'' region with Co and the
C atoms inside the circle are treated by AFQMC, with the ``outer'' region providing a frozen 
environment.}
\label{fig:Co-graphene}
\end{figure}

Figure~\ref{fig:Co-graphene} illustrates an application of AFQMC to the adsorption of  Co atoms 
on graphene. The goal of the study was to determine the stability and 
magnetic state of the Co adatom as a function of its distance from the graphene sheet. The sensitivity 
and complexity of the energetics requires a correlated treatment. In addition to serving as a useful benchmark, 
the computed results provided an explanation for experimental results with Co on free-standing 
graphene \citep{Co-graphene-Yudis-PRL}.
The AFQMC calculation was performed 
by embedding it in 
a DFT calculation to extend length scales, as illustrated on the right.
After the DFT  is 
performed, a many-body Hamiltonian is generated, using a procedure similar to that 
of producing a frozen-core Hamiltonian \citep{Wirawan-FC-JCTC13}, except the ``core'' here is actually  
the ``outer'' region  indicated on the right. (Orbital localization procedures are applied as needed.)
The resulting Hamiltonian, which describes the region inside
the shaded circle embedded in  the environment of the outer region whose orbitals are frozen, is then treated by AFQMC. 
From a QC perspective,  this approach can also be viewed as casting AFQMC as a general ``solver' for a 
(very large) active space \citep{WIRES-review}.

\section{Summary and outlook}

In this chapter, we have described a general computational framework for many-body calculations
which combines a field-theoretic description with stochastic sampling. The approach, 
referred to as auxiliary-field quantum Monte Carlo (AFQMC), is 
based on a stochastic superposition of DFT-like calculations. We have shown how the framework
can be applied to carry out \emph{ab initio} electronic structure calculations.
As mentioned, some additional references for further details include a set of lecture notes \citep{Lecture-notes} (on which some of the sections 
in this chapter are based), a pedagogical code for lattice 
models written in Matlab \citep{Matlab-code}, and 
a review on molecular systems \citep{WIRES-review}. 
 
The AFQMC approach has been applied in both condensed matter physics and quantum chemistry. 
It has been implemented with both plane-waves/pseudopotentials and with Gaussian basis sets. 
We have discussed both types of calculations, as well as a combination which uses Kohn-Sham 
orbitals generated from planewave DFT as a basis to downfold the Hamiltonian for a solid. In all of 
these, as well as in many applications to lattice models for strong electron correlation and for 
ultracold atom systems, AFQMC has shown strong promise with its scalability (with system size 
and with parallel computing platforms),
capability (total energy computation and beyond), and accuracy. 

The AFQMC method has low-polynomial (cubic) scaling with system size, by using Monte Carlo 
sampling to treat the exponential growth of the Hilbert space. It samples the many-body ground
state by a linear combination of non-orthogonal Slater determinants. The connection with independent-electron 
calculations, as we have highlighted, makes it straightforward to build AFQMC as a framework on 
top of traditional DFT or HF calculations, and take advantage of the many existing technical 
machineries developed over the past few decades in materials modeling. 

Recent developments in the computation of atomic forces and geometry optimization, and
the treatment of spin-orbit coupling and  general magnetic order, are manifestations of 
this connection. They significantly enhance the capability of stochastic methods for electronic 
structure. Similarly, the formulation for superconducting Hamiltonians and for embedding AFQMC in independent-electron calculations to extend length scales will broaden the reach 
in materials computation.

The AFQMC is approximate, because of the constraint to control the sign/phase problem. 
A major focus during the development of the framework has been to systematically test (and improve) 
the accuracy of AFQMC. A large database has now been accumulated, 
thanks in part to the major many-electron benchmark initiatives recently. 
The accuracy that can  be achieved by AFQMC with its present stage of development 
is such that many applications are now within reach in materials modeling. A variety 
of new developments are possible and currently being pursued.

The structure of the open-ended random walk, as illustrated in  Fig.~\ref{fig_OpenRW}, makes 
AFQMC ideally suited for modern high-performance computing platforms, with exceptional 
capacity for parallel scaling. The rapid growth of high-performance computing resources 
will thus provide a strong boost to the application of AFQMC in the study of molecules and 
solids. 

The development of AFQMC is entering an exciting new phase. A large number of possible directions
can be pursued, including many opportunities for algorithmic improvements and speedups. These
will be spurred forward and stimulated by growth in applications, which we hope will in turn allow more
rapid realization of a general many-body computational  framework for materials.

\begin{acknowledgement}

I  thank the many colleagues and outstanding students and postdocs whose
contributions to the work discussed here are invaluable, among whom I would especially like to 
mention
W.~Al-Saidi, H.~Krakauer, F.~Ma, M.~Motta, W.~Purwanto, and H.~Shi.
Support from the National Science
Foundation (NSF), 
the Simons Foundation, and  the Department of Energy (DOE)
is gratefully acknowledged. Computing was done via XSEDE supported by NSF,
on the Oak Ridge Leadership
Computing Facilities, and on the HPC faciliities at William \& Mary.

\end{acknowledgement}
\bibliographystyle{spbasic}  
\bibliography{HMM2-Zhang-clean} 

\end{document}